\documentclass[superscriptaddress, aps, preprintnumbers,
amsmath, amssymb, sort&compress, nofootinbib, 10pt, prl, letterpaper, floatfix]{revtex4-2}
\usepackage{amsfonts,amsmath,amssymb,ascmac,bm,tensor, microtype}
\usepackage{fnpct} 
\usepackage{comment}
\usepackage{ifpdf}
\usepackage{slashed}
\usepackage{color}
\usepackage[mathscr]{eucal}
\usepackage[utf8]{inputenc}
\usepackage{cancel}

\ifpdf
  \usepackage{graphicx}     %   usepackage without driver option
  \usepackage[bookmarksopen,colorlinks=true,linkcolor=bblue,citecolor=bblue,urlcolor=ppink]{hyperref}
\else     % For (p)LaTeX + dvipdfmx
\fi

\definecolor{red}{rgb}{1,0,0}
\definecolor{darkred}{rgb}{0.6,0,0}
\definecolor{darkgreen}{rgb}{0.992447,0.623778,0.034597}
\definecolor{ppink}{rgb}{1,0.4,0.4}
\definecolor{bblue}{rgb}{0.284602,0.317763,0.963947}
\definecolor{purple}{rgb}{0.5 ,0, 0.7}

\newcommand{\vev}[1]{ \left< {#1} \right> }
\newcommand{\dd}{\mathrm{d}}

\newcommand{\GW}{\text{GW}}
\newcommand{\GeV}{\text{GeV}}
\newcommand{\MeV}{\text{MeV}}
\newcommand{\tmax}{\text{max}}
\newcommand{\tot}{\text{tot}}
\newcommand{\ee}{\text{e}}

\newcommand{\eqf}{{\text{eq},1}}
\newcommand{\eqs}{{\text{eq},2}}

\newcommand{\fo}{{(1)}}
\newcommand{\so}{{(2)}}
\newcommand{\tho}{{(3)}}

\newcommand{\NL}{{\text{NL}}}

\newcommand{\rr}{\text{r}}
\newcommand{\kk}{\text{kin}}
\newcommand{\stf}{{*,1}}

\makeatletter
\newcommand\footnoteref[1]{\protected@xdef\@thefnmark{\ref{#1}}\@footnotemark}
\makeatother

\allowdisplaybreaks[1]

\begin{document}
 
\title{
Axion Poltergeist
}
 
\author{Keisuke Harigaya}
\affiliation{Department of Physics, University of Chicago, Chicago, IL 60637, USA}
\affiliation{Kavli Institute for Cosmological Physics and Enrico Fermi Institute, University of Chicago, Chicago, IL 60637, USA}
\affiliation{Kavli Institute for the Physics and Mathematics of the Universe (WPI),
The University of Tokyo Institutes for Advanced Study,
The University of Tokyo, Kashiwa, Chiba 277-8583, Japan}

\author{Keisuke Inomata}
\affiliation{Kavli Institute for Cosmological Physics and Enrico Fermi Institute, University of Chicago, Chicago, IL 60637, USA}

\author{Takahiro Terada}
\affiliation{Particle Theory and Cosmology Group, Center for Theoretical Physics of the Universe, 
Institute for Basic Science (IBS), Daejeon, 34126, Korea}

\preprint{CTPU-PTC-23-19}

\begin{abstract}
\noindent
Rotations of axion fields in the early universe can produce dark matter and the matter-antimatter asymmetry of the universe.  We point out that the rotation can generate an observable amount of a stochastic gravitational-wave (GW) background.  It can be doubly enhanced in a class of models in which the equation of state of the rotations rapidly changes from a non-relativistic matter-like one to a kination-like one  by 1) the so-called poltergeist mechanism and 2) slower redshift of GWs compared to the axion kination fluid. In supersymmetric UV completion, future GW observations can probe the supersymmetry-breaking scale up to $10^7$\,GeV even if the axion does not directly couple to the Standard Model fields.
\end{abstract}

\maketitle

\twocolumngrid

{\it Introduction.}
Gravitational waves (GWs) are powerful probes of the early universe because all the matter in the universe can in principle produce GWs through gravitational interactions and the produced GWs are not attenuated except by cosmic expansion. 
For example, GWs can probe first-order phase transitions associated with spontaneous symmetry breaking (SSB) in the early universe~\cite{Caprini:2015zlo,Caprini:2018mtu,Caprini:2019egz,LISACosmologyWorkingGroup:2022jok}.
In this letter, we shed light on GWs induced by the dynamics of a Nambu-Goldstone boson arising from SSB.

Global symmetry and its SSB play important roles in solutions to the problems in the Standard Model of particle physics, including 
the strong CP problem~\cite{Peccei:1977hh,Peccei:1977ur}, non-zero neutrino masses~\cite{Chikashige:1980ui}, and flavor hierarchies~\cite{Froggatt:1978nt}.
An important prediction of SSB is a Nambu-Goldstone boson~\cite{Nambu:1961tp,Goldstone:1961eq,Goldstone:1962es}.
If the symmetry is only approximate and is violated by a small amount, the boson obtains a small mass and is called a pseudo Nambu-Goldstone boson. 
In particular, the boson predicted in Peccei-Quinn's solution to the strong CP problem is called the (QCD) axion~\cite{Weinberg:1977ma, Wilczek:1977pj}. Others are often called axion-like particles (ALPs), but in this letter, we simply call any pseudo Nambu-Goldstone bosons associated with SSB of U(1) symmetry as axions. Axions are naturally light and weakly coupled to other fields, and hence are sufficiently long-lived to be dark matter (DM) of the universe~\cite{Preskill:1982cy,Abbott:1982af,Dine:1982ah}. Axion searches are performed and planned in a wide range of its mass and couplings to the Standard Model particles~\cite{Adams:2022pbo}.

Because of the lightness, axions can exhibit interesting dynamics in the early universe and
play cosmological roles.
The most commonly considered dynamics is oscillation, which may explain the observed DM abundance~\cite{Preskill:1982cy,Abbott:1982af,Dine:1982ah}.
In this work, we instead consider rotation of an axion in the field space, which can be initiated by the Affleck-Dine mechanism~\cite{Affleck:1984fy} and has rich cosmological, astrophysical, and particle-physics implications~\cite{Kamada:2019uxp,Co:2019jts,Co:2020xlh,Co:2022qpr,Eroncel:2022efc}.
The rotation can produce axion DM via kinetic misalignment~\cite{Co:2019jts,Co:2020dya,Co:2021rhi,Eroncel:2022vjg}, which predicts larger couplings of the axion
to Standard model particles than what the conventional production mechanisms predict. The produced axion may have large density fluctuations and form mini halos~\cite{Eroncel:2022efc}, which can be observed through gravitational lensing~\cite{Arvanitaki:2019rax}. The rotation can also produce the matter-antimatter asymmetry of the universe~\cite{Co:2019wyp}. Simultaneous production of DM and the matter-antimatter asymmetry from axion rotation strongly constrains the axion parameter space and/or predicts signals in particle-physics experiments~\cite{Co:2020xlh,Co:2020jtv,Harigaya:2021txz,Chakraborty:2021fkp,Kawamura:2021xpu,Co:2021qgl,Barnes:2022ren,Co:2022kul,Badziak:2023fsc}.

GW signals of the axion rotation have been discussed in the literature. Axion rotation can amplify GWs produced by quantum fluctuations during inflation or by cosmic strings~\cite{Co:2021lkc,Gouttenoire:2021wzu,Gouttenoire:2021jhk}. An axion-dark photon coupling can also produce GWs~\cite{Co:2021rhi,Madge:2021abk}.

In this letter, we point out that the rotation itself can generate a substantial stochastic GW background through its gravitational interaction at the second order in perturbations.
For a certain potential of the symmetry-breaking field, the axion rotation initially follows the equation of state of non-relativistic matter and dominates the universe. 
At some point, the equation of state suddenly changes to that of kinetic-energy-dominated fluid and, after a while, the energy density of the rotation eventually becomes subdominant compared with that of radiation~\cite{Co:2019wyp}. 
The sudden transition from a matter-dominated (MD) era to a kinetic-energy-dominated (KD or kination~\cite{Spokoiny:1993kt, Joyce:1996cp, Ferreira:1997hj}) era can produce strong GWs through rapid oscillations of the density perturbations with nonzero sound speed after the transition.
This ``poltergeist mechanism''~\cite{Inomata:2019ivs, Inomata:2020lmk}, where a ``ghost" of non-relativistic matter produces sounds that generate GWs, has been studied for the sudden transition from a MD era to a radiation-dominated (RD) era~\cite{Inomata:2019ivs, Inomata:2020lmk, White:2021hwi, Domenech:2020ssp, Domenech:2021wkk, Lozanov:2022yoy,Bhaumik:2022pil,Bhaumik:2022zdd,Kasuya:2022cko,Domenech:2023mqk}.
GWs induced during an era with a general equation of state have been discussed in Refs.~\cite{Domenech:2019quo, Domenech:2020kqm}.

Throughout this letter, we take the natural unit $c= \hbar = 8\pi G = k_\text{B} = 1$. We take the conformal Newtonian gauge\footnote{Though our calculation is performed in the Newtonian gauge, the final result is the gauge independent physical contribution.  This point is explained in SM in connection to the gauge dependence issue discussed in the literature.}, in which the nonvanishing components of the metric are
\begin{align}
    g_{00} =& -a^2(1+2\Phi), &
    g_{ij} =& a^2 \! \left((1 - 2 \Psi)\delta_{ij} +\frac{h_{ij}}{2} \right) \! ,
    \label{eq:newton_gauge}
\end{align}
where $a$ is the scale factor, the gravitational potential $\Phi$ and the curvature perturbation $\Psi$ are the first-order scalar perturbations, the GWs $h_{ij}$ are the second-order tensor perturbations. We have omitted the vector perturbations and the first-order tensor perturbations because they are irrelevant to the GW production mechanism in this letter.
Since we focus on the evolution of the universe dominated by scalar fields, where the anisotropic stress vanishes~\cite{Weinberg:2008zzc}, we set the perfect fluid condition, $\Phi = \Psi$, in the following.

\vspace{0.2cm}
{\it Matter-to-kination transition and gravitational waves.}
We here explain the essence of the poltergeist mechanism.
The tensor perturbations in the Fourier space are induced by the scalar perturbations through~\cite{Ananda:2006af, Baumann:2007zm}
\begin{align}
{h^{\lambda}_{\bm{k}}}''\!(\eta) + 2 \mathcal H {h^{\lambda}_{\bm{k}}}'\!(\eta) + k^2 h^{\lambda}_{\bm{k}}(\eta) = 4 S^{\lambda}_{\bm{k}}(\eta),
\label{eq:h_eom}
\end{align}
where $\bm{k}$ ($k = |\bm{k}|$) and $\lambda$ denote the wavenumber and the polarization of the tensor perturbations, the prime is $\partial/\partial \eta$, and $\mathcal H = a'/a$. % = a H$. 
See also Supplemental Material (SM) for the detailed definition of the Fourier mode, $h^\lambda_{\bm k}$.
The source term $S^\lambda_{\bm k}$ is given by
\begin{align}
\label{eq:s}
S^\lambda_{\bm{k}} =&\,  \int \! \frac{\dd^3 q}{(2\pi)^3} \, e^{\lambda}_{ij}(\hat{\bm k})q^i q^j \bigg[ 2 \Phi_{\bm{q}} \Phi_{\bm{k-q}}  
 + \frac{4  \widehat{\Phi}_{\bm{q}} \widehat{\Phi}_{\bm{k-q}}}{3(1+w)}  \bigg],
\end{align}
where $\widehat \Phi_{\bm k} \equiv \mathcal{H}^{-1}\Phi'_{\bm{k}} + \Phi_{\bm{k}}$, $\hat k = \bm k/k$, $e^\lambda_{ij}(\hat{\bm k})$ is the polarization tensor, and $w = p/\rho$ is the equation-of-state parameter with $p$ and $\rho$ the pressure and the energy density, respectively.
Also, for one-component fluid (no entropy perturbations), the equation of motion of $\Phi$ is given by~\cite{Mukhanov:991646}
\begin{align}
        &\Phi_{\bm{k}}'' + 3 \! \left(1+c_s^2 \right) \! \mathcal H \Phi_{\bm{k}}' + \! \left( c_s^2 k^2 \! + 3 \! \left(c_s^2-w \right) \! \mathcal H^2 \right) \! \Phi_{\bm{k}} = 0,
        \label{eq:phi_eom}
\end{align}
where $c_s$ is the sound speed.
The amplitude and the evolution of the gravitational potential $\Phi$ determine the amount of the induced GWs.

In this work, we numerically follow the evolution of $\Phi$ from a time in the MD (axion-rotation-dominated) era by solving Eq.~(\ref{eq:phi_eom}) with the initial condition $\Phi = -3M\zeta/5$ and $\Phi'= 0$ even for the perturbations that enter the horizon before the MD era.
The evolution of the gravitational potential before the initial time is incorporated in the factor $M$ that  depends on the scale. 
See SM for the concrete expression of $M$.
Note that this initial condition is based on the assumption that the isocurvature perturbations of the axion rotation are negligible compared to the curvature perturbations.

In the poltergeist mechanism, the subhorizon perturbations that enter the horizon before the sudden transition play important roles after the transition.
The gravitational potential on subhorizon scales is constant during the MD era due to the growth of the density perturbations. 
Once the universe enters the era with $c_s^2 \neq 0$, the subhorizon gravitational potential starts to oscillate with the timescale $1/(c_s k)$, which can be much shorter than the Hubble timescale $\eta$.
If the transition occurs suddenly, the amplitude of the gravitational potential is not much suppressed before the KD era begins.
The terms $\mathcal H^{-1} \Phi'_{\bm k} (\sim \mathcal O(k\eta) \Phi_{\bm k})$ in Eq.~(\ref{eq:s}) source GWs the most
because of the large factor $k\eta \gg 1$.
Figure~\ref{fig:phi} shows the evolution of the subhorizon gravitational potential in a concrete model that we will explain below.
The case with $d = 5\times 10^{-5}$ approximates the instantaneous transition limit, though the case with $d = 0.05$ still leads to strong GWs, as we will see below.

In addition to the poltergeist mechanism, the energy density parameter of the induced GWs is further enhanced because GWs are produced from the kination fluid that has a larger energy density than the radiation and later becomes subdominant without producing entropy.

\begin{figure}
        \centering \includegraphics[width=0.99\columnwidth]{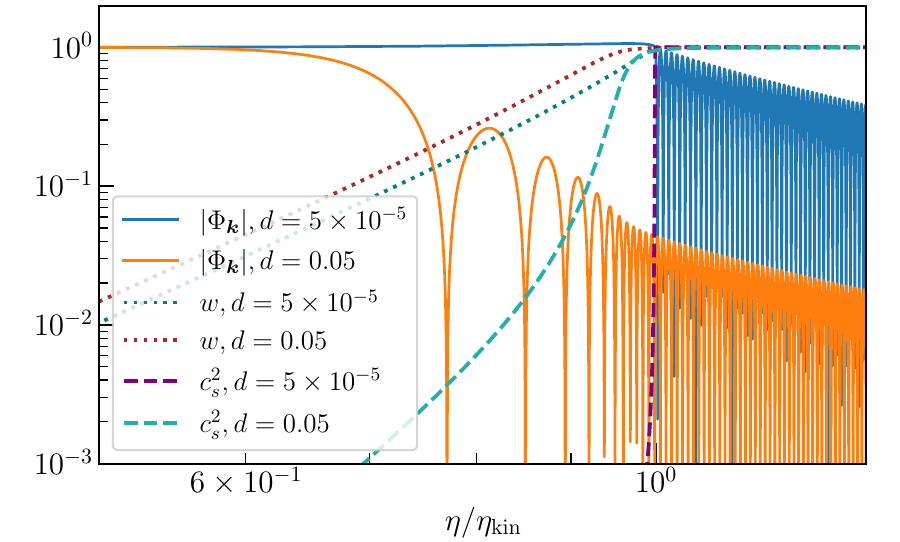}
        \caption{ The evolution of the gravitational potential in the two-field model, whose Lagrangian is given by Eq.~(\ref{eq:Leff}).
        The gravitational potential is normalized so that $\Phi_{\bm k} = 1$ during the MD era.
        We take $k \eta_\kk = 450$ for $\Phi_{\bm k}$, which is close to the nonlinear density perturbation scale during the MD era with $\mathcal P_\zeta = 2.1\times 10^{-9}$~\cite{Assadullahi:2009nf,Kohri:2018awv,Inomata:2019zqy} (see also SM).
        }
        \label{fig:phi}
\end{figure}

\vspace{0.2cm}
{\it Axion rotation.}
A sudden change of the equation of state indeed occurs in the rotational dynamics of an axion.
The axion is the angular direction $\theta$ of a complex scalar field $P$ that has a nearly $\mathrm{U}(1)$-symmetric wine-bottle potential. If the potential is flat, as naturally occurs in supersymmetric theories, it may take on a large field value in the early universe. We further assume that the $\mathrm{U}(1)$ symmetry is explicitly broken by a higher dimensional operator in the potential of $P$. The explicit $\mathrm{U}(1)$ breaking becomes effective for a large field value of $P$ and drives angular motion~\cite{Affleck:1984fy}. Because of the cosmic expansion, the radial field value of $P$ decreases and the higher-dimensional operator soon becomes ineffective. $P$ continues to rotate while preserving its angular momentum in the field space, i.e., a $\mathrm{U}(1)$ charge. We call this motion ``axion rotation".

The rotation is initially elliptic and superposition of angular motion and oscillating radial motion. 
The latter is dissipated via the interaction with the thermal bath. 
On the other hand, if the $\mathrm{U}(1)$ charge density is larger than $m_S T^2$ with $T$ being the temperature of the thermal bath and $m_S$ being the mass of the radial direction, the angular motion remains almost intact, since that is the state with the minimal free energy for a fixed $\mathrm{U}(1)$ charge~\cite{Laine:1998rg,Co:2019wyp,Domcke:2022wpb}.

We consider the case where the potential of $P$ is nearly quadratic at large field values. The energy density of the universe evolves as follows~\cite{Co:2019wyp}. When the axion rotates at the body of the potential, the energy density decreases in proportion to $a^{-3}$ while the radius of the rotation shrinks. Even if the universe is initially dominated by radiation, because of the matter-like behavior, the axion rotation can dominate the universe at $\eta=\eta_\eqf$, which realizes the MD era. The axion reaches the bottom of the potential at $\eta=\eta_\kk$, after which the energy density decreases in proportion to $a^{-6}$; the equation of state is that of kination.
The axion rotation eventually becomes subdominant again at $\eta=\eta_\eqs$ and the second (standard) RD era begins.

A concrete setup is given by a supersymmetric two-field model whose effective Lagrangian is given by~\cite{Co:2021lkc}
\begin{align}
\label{eq:Leff}
    {\cal L} &=  \! \left( 1 + \frac{f_a^4}{16|P|^4} \right)  |\partial P|^2 -  m^2_S \! \left( |P|  - \frac{(1+d)f_a^2}{4|P|}\right)^{\! 2} \! ,
\end{align}
where $|\partial P|^2 \equiv - g^{\mu\nu}\partial_\mu P^\dag \partial_\nu P$, $m_S$ is the mass parameter of the radial mode given by supersymmetry breaking, $f_a$ is the axion decay constant, and $1+d$ is a ratio of mass parameters in the UV completion of the model. 
We assume $d \geq 0$; for $d < 0$, the axion rotation is unstable~\cite{Co:2021lkc}.
The perturbativity of the model requires $f_a > m_S$.
See SM for the detail.
For $|P|\gg f_a$, the potential is indeed nearly quadratic.

$\eta_\eqs$ and $\eta_\kk/\eta_\eqf$ are related to the model parameters in Eq.~(\ref{eq:Leff}) and the charge density normalized by the entropy density $s$, which we denote by $Y_\theta = 2 \left( 1 + f_a^4/(16|P|^4) \right) |P|^2\theta'/(a s)$~\cite{Co:2021lkc}.
For $\eta_\eqs$, we have
\begin{align}
    \frac{1}{2\pi \eta_\eqs} =1.1\times 10^{-5}\,\text{Hz} \times \left( \frac{T_\eqs}{1.8\times 10^2\,\GeV} \right),
\end{align}
where $T_\eqs$ is the temperature at $\eta_\eqs$,
\begin{align}
            T_\eqs &= 1.8\times 10^2\, \GeV \times \left( \frac{f_a}{10^6\, \GeV} \right) \left( \frac{Y_\theta}{10^3}\right)^{-1}.
            \label{eq:t_eqs}
\end{align}
Consistency with the big-bang nucleosynthesis (BBN) prediction requires $T_\eqs \gtrsim 2.5\,\MeV$~\cite{Co:2021lkc}.
$\eta_\kk/\eta_\eqf$ is 
\begin{align}
            \frac{\eta_\kk}{\eta_\eqf} = 
            1.7\times10^2  \left(  \frac{m_S}{10^5\, \GeV} \! \right)^{\!\frac{1}{3}} \!\left( \! \frac{f_a}{10^6\, \GeV} \! \right)^{\! \! -\frac{1}{3}}\! \! \left( \! \frac{Y_\theta}{10^3}\!\right)^{\!\frac{2}{3}}\!\!.
\end{align}

$Y_\theta$ is maximized when the initial elliptic rotation has $\mathcal O(1)$ ellipticity, dominates the universe, and then gets thermalized. The maximal value of $Y_\theta$ is~\cite{Co:2021lkc}
\begin{align}
            Y_{\theta,\tmax} = 10^3 \left(\frac{m_S}{8.7 \times 10^5\, \GeV}\right)^{-\frac{1}{3}} \left( \frac{b}{0.1}\right)^{\frac{1}{3}},
\end{align}
where we have used Eq.~(\ref{eq:t_eqs}) and $b \lesssim0.1$ is a thermalization model-dependent constant.

If axion DM is produced by the rotation via the kinetic misalignment, its mass $m_a$ is related to $Y_\theta$ as $C m_a Y_\theta \simeq 0.44\, \text{eV}$~\cite{Co:2021lkc}, where $C$ is a constant that is expected to be $\mathcal O(1)$ and its exact value is not yet determined. In this work, we take $C=1$ as a reference value.

Since $P$ is a complex field, the perturbations of $P$ in general have two modes. However, after the oscillating radial mode is dissipated, in the limit where the rotation is rapid, $m_S\gg \mathcal H/a, k/a$, one mode can be integrated out so that the perturbations are effectively those of one-component (adiabatic) fluid~\cite{Co:2021lkc}.
The equation-of-state parameter $w$ of the axion rotation in this model is derived in Ref.~\cite{Co:2021lkc}. We compute the sound speed of the rotation by using the adiabatic sound speed $c_s^2 = p'/\rho'= w - w'/(3 \mathcal H(1+w))$, which is shown in Fig.~\ref{fig:phi}.
See also SM for the derivation of this adiabatic relation from the definition of the sound speed ($c_s^2 \equiv \delta p/\delta \rho$) for a rotating complex field.
We follow the evolution of the energy density and the pressure by using the field equation and the expression of the energy momentum tensor (see SM and Ref.~\cite{Co:2021lkc}).
In Fig.~\ref{fig:phi}, one can see that, for $d\ll1$, the sound speed changes rapidly around the matter-to-kination transition. 
This rapid change of the sound speed is the origin of the poltergeist mechanism. $d\ll1$ is natural when the UV theory enjoys approximate $Z_2$ symmetry.
For convenience, we define $\eta_\kk$ as the time when $c_s^2 = 0.95$ and use this as the beginning of the KD era throughout this work.
Note that the evolution of the normalized gravitational potential as a function of $\eta/\eta_\kk$ depends only on $d$ and the wavenumber $k\eta_\kk$.

\vspace{0.2cm}
{\it Setup.}
We here explain our fiducial setup.
We consider a scale-invariant power spectrum for the curvature perturbations with the cutoff scales: 
\begin{align}
	\mathcal P_\zeta = A\, \Theta(k_\tmax-k)\Theta(k-1/\eta_\kk).
	\label{eq:pz}
\end{align}
We introduce an IR cutoff $1/\eta_\kk$ to reduce the computational cost by focusing on the contributions from the poltergeist mechanism, which comes from the scalar perturbations that enter the horizon during the MD era. 
In our fiducial parameter sets, the contributions from $k < 1/\eta_\kk$ are subdominant compared to the poltergeist contribution except on the scales much larger than the peak scale.
We also introduce a UV cutoff $k_\tmax$ to obtain a conservative GW spectrum within the linear perturbation theory.
Smaller-scale density perturbations enter the horizon earlier and finally become non-perturbative earlier.
We take the same amplitude as the CMB fluctuations, $A = 2.1\times 10^{-9}$, as a conservative reference value.
If the axion isocurvature perturbations are not negligible or the adiabatic perturbation is highly blue-tilted, $A$ can be larger.

For simplicity, we assume that the effective degrees of freedom is constant with $g_* = 106.75$ in $\eta_\eqf < \eta < \eta_\eqs$.
Also, due to the difficulty of the numerical calculation, we only take into account the GWs induced after the KD era begins ($\eta > \eta_\kk$), which gives conservative results.

We here summarize the limitation of the linear perturbation theory, which our analysis is based on. 
First, the perturbations of the axion field may become nonlinear because the sudden change of the state of the universe, characterized by $w$ and $c_s^2$, enhances higher-order perturbation contributions.
Similar enhancements of the higher-order contributions have been studied in the inflation models with sharp features, which cause the sudden change of the slow-roll parameters~\cite{Adshead:2014sga,Inomata:2021uqj,Inomata:2021tpx}.
We have numerically confirmed that the case of $d=0.05$ marginally remains in the linear regime, but the case of $d=0.01$ marginally does not.
Because of this, we take $d=0.05$ as a benchmark value. 
See SM for the technical details of the evaluation of the nonlinearity.
In this sense, the case of $d =5\times 10^{-5}$ in Fig.~\ref{fig:phi} is unreliable.
Nevertheless, we plot this case in Fig.~\ref{fig:phi} to show the essential idea of the sharp transition.

Second, the density perturbations
that enter the horizon long before the second RD era may become non-perturbative.
This is because the subhorizon density perturbation of the axion rotation, $\delta$, grows as $\propto \ln \eta$ during the first RD era, $\propto \eta^2$ during the MD era, and $\propto \eta^{1/2}$ during the KD era, up to its oscillations.
Since this growth begins after the horizon entry of the modes, the perturbations on the smallest scale $k_\tmax$ first reaches $|\delta| = 1$ if the MD era lasts long enough.
If the density perturbation becomes $|\delta| > 1$, the cosmological perturbation theory is no longer reliable.
On the other hand, once the induced GWs enter the horizon, they are decoupled from the source terms and freely propagating~\cite{Domenech:2020kqm}.
Owing to this, we can safely calculate the induced GWs on the scales that enter the horizon before the smallest-scale perturbations become $|\delta| > 1$. 
Technical details on our choice of $k_\tmax$ can be seen in SM.

\vspace{0.2cm}
{\it Results.}
Figure~\ref{fig:gw} shows the induced GW spectra in our setup. 
We can see that the induced GWs can be probed by the future observations.
The peak scale is around
$k_* \equiv \text{min}[k_\tmax,1/\eta_\eqf]$. 
The analytical estimate of the GW spectrum in $k<k_*$ is given by 
\begin{align}
        \Omega_\GW h^2
        &\simeq 2\times 10^{-11} A^2 Q^4 B(k) \, \frac{\eta_\kk^2}{\eta_\eqf^2} \,  k_*^5 k \eta_\kk^{6},
        \label{eq:omega_gw_app}
\end{align}
where $Q$ corresponds to the amplitude of the normalized $\Phi$ on $k_*$ at $\eta_\kk$ (up to the oscillations), and $B(k)$ is $1$ for $k>4/\eta_\kk$ and $0.535 \times(k\eta_\kk)$ for $k < 4/\eta_\kk$. 
The factor $\eta_\kk^2/\eta_\eqf^2 >1$ in Eq.~(\ref{eq:omega_gw_app}) is a redshift factor, which comes from the fact that the energy density of the rotating axion field and the induced GW are $\propto a^{-6}$ and $a^{-4}$, respectively.
See SM for the derivation of this analytical estimate and how to numerically calculate the induced GWs, where we also extend the analyses to a general value of $w$.

\begin{figure}
        \centering \includegraphics[width=0.99\columnwidth]{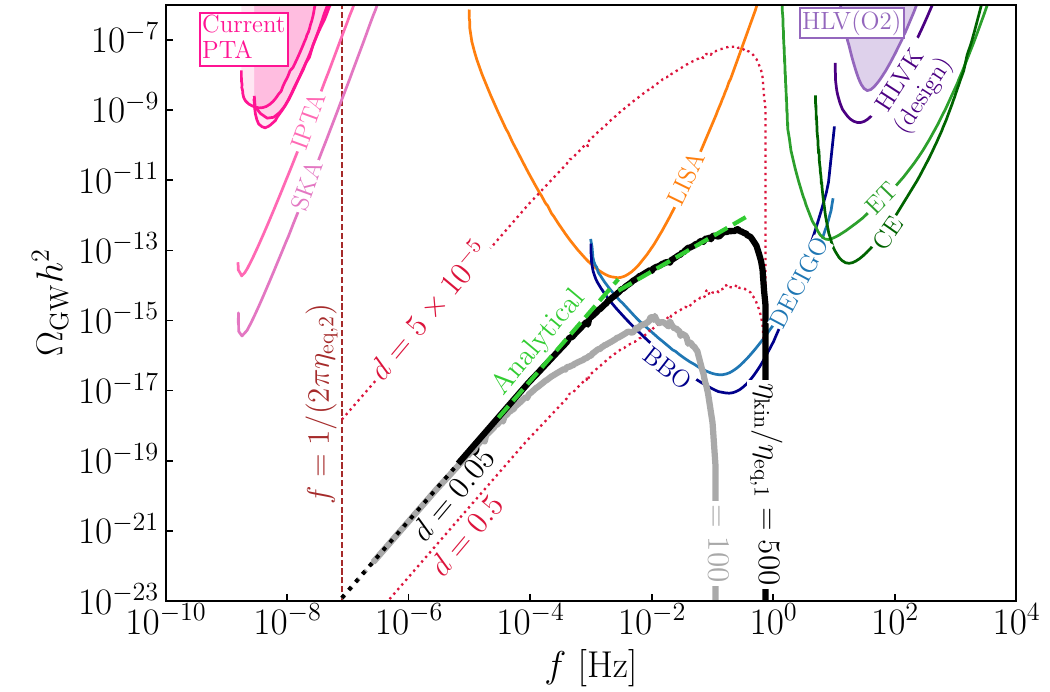} 
        \caption{ 
        The induced GW spectrum and the sensitivities of the future and current experiments~\cite{Schmitz:2020syl}.
        The shaded regions are excluded by the current observations. 
        The black and gray lines are the GW spectrum with $d=0.05$, $\eta_\eqs = 2\times 10^6\,$s, and different $\eta_\kk/\eta_\eqf$.
        The curvature power spectrum is given by Eq.~(\ref{eq:pz}).
        The green dashed line shows the analytical estimate, Eq.~(\ref{eq:omega_gw_app}), for the black-line case.
        The black and gray dotted lines show the region $f < 1/(2\pi \times 100 \eta_\kk)$, which are superhorizon modes when the smallest-scale perturbations become nonlinear, $\delta \simeq \mathcal O(1)$.                
        For comparison, the spectra for different values of $d$ with $\eta_\kk/\eta_\eqf = 500$ are shown with the red dotted lines.
        }
        \label{fig:gw}
\end{figure}

Figure~\ref{fig:ms_fa} shows the regions in the $(m_S,f_a)$ plane that can be probed by the future GW experiments for given $Y_\theta/Y_{\theta,\tmax}$ with signal-to-noise ratio (SNR) $>1$.
See SM for how to calculate the SNR for each observation.
We can see that the future observations can investigate $\mathcal O(10^{-1})\,\GeV \lesssim m_S \lesssim \mathcal O(10^7)\,\GeV$ and $\mathcal O(10^{2})\,\GeV \lesssim f_a \lesssim \mathcal O(10^8)\,\GeV$, depending on the value of $Y_\theta/Y_{\theta,\tmax}$.
Note that these parameter regions are testable even if the axion is not coupled to the photon.

\begin{figure}%[t] 
        \centering \includegraphics[width=0.99\columnwidth]{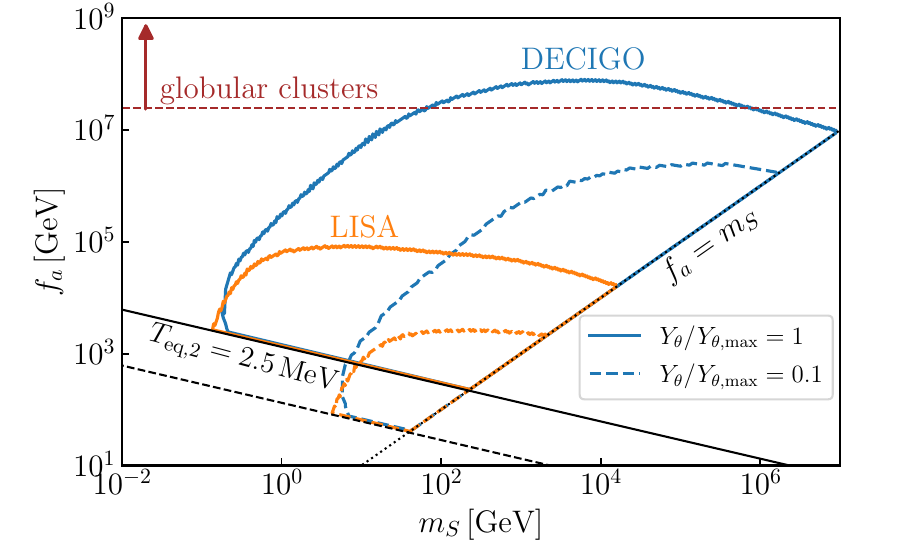}
        \caption{         
        The parameter regions where LISA~\cite{LISACosmologyWorkingGroup:2022jok} and DECIGO~\cite{Kawamura:2020pcg} can probe for fixed $Y_\theta/Y_{\theta,\tmax}$.
        The region surrounded by the lines realizes the SNR $>1$ for each project with 1-year observation.
        $d=0.05$ and $b = 0.1$ are taken for all lines.
        The boundaries of the BBN constraint $T_\eqs > 2.5 \,\MeV$, which depends on $Y_\theta$, and the perturbativity constraint $f_a > m_S$ are shown by black lines.
        BBO has a similar sensitivity as DECIGO.
        Other projects in Fig.~\ref{fig:gw} do not have a region with $\text{SNR} > 1$.
        The parameter region below the horizontal brown line is constrained from globular clusters~\cite{Dolan:2022kul} if the coupling between axion and photon is given by $g_{a\gamma\gamma} = \alpha/(2\pi f_a)$ with $\alpha$ the fine structure constant.
        }
        \label{fig:ms_fa}
\end{figure}

Figure~\ref{fig:ma_fa} shows the observable regions in the $(m_a,f_a)$ plane for given $m_S$ when the axion DM is produced from the rotation.
The horizontal and vertical cuts in the observable regions come from the constraints $f_a > m_S$ and $Y_\theta < Y_{\theta,\tmax}$, respectively. 
The black dot-dashed line is the prediction of the QCD axion and 
the region below the dark-cyan dot-dashed line is the prediction of the scenario where the matter-antimatter asymmetry of the universe is generated by the axion rotation and the electroweak sphaleron process, called ALP cogenesis~\cite{Co:2020xlh}.

\begin{figure}%[tbhp!] 
        \centering \includegraphics[width=0.99\columnwidth]{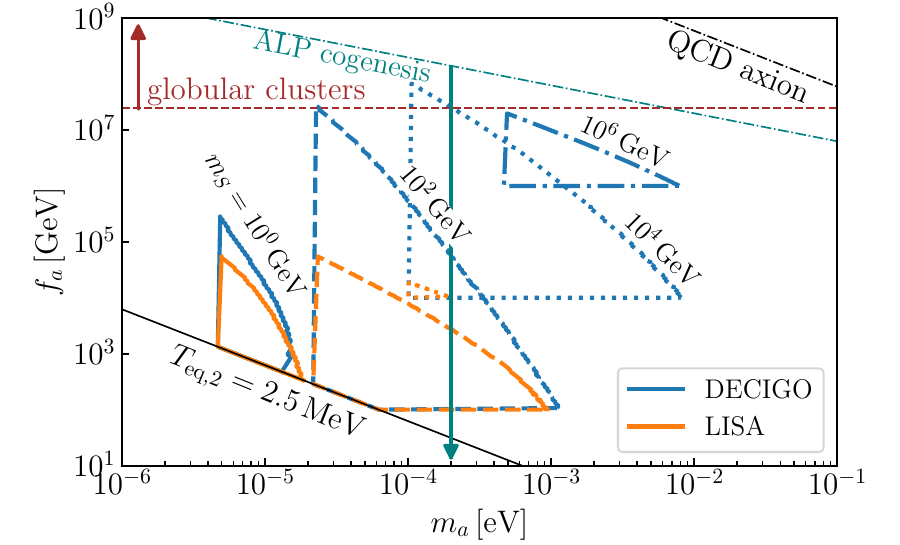}
        \caption{ 
        The regions in the $m_a$-$f_a$ plane with fixed $m_S$ that can be probed by future GW observations with SNR $>1$ when axion DM is produced from the rotation. 
        The prediction of the QCD axion and the ALP cogenesis are also shown.
        For the ALP cogenesis, the region below the line is the prediction. 
        See the caption of Fig.~\ref{fig:ms_fa} for the explanation on the other lines.
        }
        \label{fig:ma_fa}
\end{figure}

\vspace{0.2cm}
{\it Summary and discussion.}
In this letter, we have pointed out that axion rotation can produce strong GWs by the poltergeist mechanism through the sudden transition from a MD era to a KD era.
The produced GWs may be abundant enough to be detected by future GW observations.
These GW signals do not rely on the coupling of the axion to the Standard Model particles and therefore enable us to investigate the uncharted parameter region of axion models. The mass of the U(1) symmetry-breaking field is given by supersymmetry breaking, and GW observations can probe supersymmetry-breaking scale as high as $10^7$ GeV.

The poltergeist mechanism in our setup is realized by the approximately homogeneous one-component fluid (i.e., the axion rotation) whose equation of state changes rapidly.  This is advantageous compared to the existing examples of the poltergeist mechanism, which are based on the simultaneous evaporation of localized objects such as black holes. In such cases, the distributions of mass and spin must be sufficiently narrow~\cite{Inomata:2020lmk}.  Our scenario is free from this issue and thus more robust against cosmological/astrophysical uncertainties.

We also mention possible extensions of this work.
1) We neglected the contributions from nonlinear density fluctuations by introducing $k_\tmax$ in this work, but they may produce non-negligible GWs. 
Since these GWs cannot be calculated within the linear perturbation theory, we need to calculate them with a non-perturbative method, such as lattice simulations and the $N$-body simulations~\cite{Kawasaki:2023rfx}.
See also Refs.~\cite{Jedamzik:2010dq, Jedamzik:2010hq} for earlier attempts.
2) When the curvature power spectrum is enhanced on small scales or the isocurvature perturbations of the axion rotation are large, an observable amount of GWs can be produced in a wider class of cosmological histories and axion models.
If the matter-to-kination transition is sudden and the axion rotation dominates for a long period,
the poltergeist mechanism dominantly produces GWs 
and our analysis in this letter is applicable just by increasing $A$, though the cutoff $k_\tmax$ becomes smaller because the density perturbations reach unity within a shorter time period after its horizon entry.
Even if the transition is not so sudden, or
the axion rotation domination occurs only for a short period or does not occur at all, strong GWs can still be induced because of the large density perturbations.
Computation of the GW spectrum for these cases, especially for the no-axion-domination case, requires careful calculation of the evolution of the density perturbations of the two fluids, axion and radiation.
3) The large axion isocurvature perturbations can also produce primordial black holes via the collapse of large fluctuations.
During the axion-dominated era, the large axion isocurvature perturbations become the curvature perturbations and PBHs can be produced from the large curvature perturbations. 
In our scenarios, the first part of the axion-dominated era is a MD era and the PBH production rate is enhanced during the MD era due to the absence of the pressure in fluid~\cite{Khlopov:1980mg,Khlopov:1982sov,Harada:2016mhb,Harada:2017fjm}.
This would result in the increase of the PBH abundance only within the mass range that corresponds to the horizon scales during the MD era.
Investigations for these topics are left for future work.

{\bf Acknowledgment.}
K.I.~and T.T.~thank Kyohei Mukaida for discussions on axion rotation physics at the initial stage of this work. K.H.~thanks Raymond Co, Nicolas Fernandez, Akshay Ghalsasi, and Jessie Shelton for collaboration on a related project.  
K.I.~was supported by the Kavli Institute
for Cosmological Physics at the University of Chicago
through an endowment from the Kavli Foundation and
its founder Fred Kavli.
K.H.~was partly supported by Grant-in-Aid for Scientific Research from the Ministry of Education, Culture, Sports, Science, and Technology (MEXT), Japan (20H01895) and by World Premier International Research Center Initiative (WPI), MEXT, Japan (Kavli IPMU).
The work of T.T.~was supported by IBS under the project code, IBS-R018-D1.

%%%%%%%%%%%%%%%%%%%%%%%%%%%%%%%%%%%
%%%%%%%%%%%%%%%%%%%%%%%%%%%%%%%%%%%
%%%%%%%%%%%%%%%%%%%%%%%%%%%%%%%%%%%
\small
\bibliographystyle{apsrev4-1}
\bibliography{axion_poltergeist}

\clearpage
\newpage
\maketitle
\onecolumngrid
\begin{center}
\textbf{\large Axion Poltergeist} \\ 
\vspace{0.05in}
{ \it \large Supplemental Material}\\ 
\vspace{0.05in}
{Keisuke Harigaya, Keisuke Inomata, and Takahiro Terada}
\end{center}
\onecolumngrid
%%%%%%%%%% Merge with Supplemental material %%%%%%%%%%
\setcounter{equation}{0}
\setcounter{figure}{0}
\setcounter{table}{0}
\setcounter{section}{0}
\setcounter{page}{1}
\makeatletter
\renewcommand{\theequation}{S\arabic{equation}}
\renewcommand{\thefigure}{S\arabic{figure}}

\setcounter{secnumdepth}{1}

This Supplemental Material is organized as follows.
In Sec.~\ref{smsec:eom}, we summarize the equations of motion for a rotating field and derive the expressions of the equation-of-state parameter and the sound speed, which are used to calculate the evolution of the gravitational potential. 
In Sec.~\ref{smsec:two_field_model}, we discuss the UV completion of the effective Lagrangian of the two-field model in the main text and the field evolution in the effective Lagrangian. 
Then, we discuss the evolution of the scale factor, the Hubble parameter, and the gravitational potential through the transitions of multiple eras in the early universe in Sec.~\ref{smsec:a_and_h}.
In Sec.~\ref{smsec:induced_gw}, we explain how to numerically and analytically calculate the GWs induced by the perturbations of the rotating axion via the poltergeist mechanism.
Finally, we discuss the limitation of the linear perturbation analysis in Sec.~\ref{smsec:limit_linear}.

\section{Equations of motion for a rotating field}
\label{smsec:eom}

In this section, we summarize the equations of motion for a rotating scalar field. Using the equations, we also relate the equation-of-state parameter $w$ and the sound speed $c_s$ to the potential of the field.
The calculation in this section generalizes the results in Ref.~\cite{Co:2021lkc} to the case with more general forms of the kinetic terms by introducing $\xi$ and $\chi$ defined below.
We consider the following Lagrangian: 
\begin{align}
        \mathcal L = -\frac{\xi(S)}{2} \partial^\mu S \partial_\mu S - \frac{\chi(S) S^2}{2} \partial^\mu \theta \partial_\mu \theta - V(S),
        \label{eq:L_gen_S}
\end{align}
where $S$ is the radial field and $\theta$ is the phase field. 
This form of Lagrangian includes the two-field model, which we discuss in the main text.
In the following, we discuss the equations of motion for the fields at the background and the linear perturbation levels. 
The equations of motion can be obtained by taking the variation of the action $\mathcal S = \int \dd^4 x \sqrt{-g} \mathcal L$, where $g$ is the determinant of the matrix $g_{\mu\nu}$.
See also Ref.~\cite{Weinberg:2008zzc} for the derivation of the equations of motion from a given Lagrangian.

At the background level, the equations of motion are 
\begin{align}
&\bar \theta'' + \left(\ln\left(a^2 \chi \bar S^2 \right)\right)' \bar \theta' = 0, \label{eq:thbar_eom_g} \\
&\bar{S}'' +2 \mathcal{H} \bar{S}'+\frac{\left(\bar{S}'\right)^2 \xi_{,\bar S}}{\xi} + \frac{1}{\xi} \left( a^2 V^\fo - \frac{1}{2}\left(\chi \,\bar{S}^2\right)_{,\bar S} \left(\bar{\theta }'\right)^2 \right)=0, \label{eq:rbar_eom_g}
\end{align}
where $V^{(n)} \equiv \partial^n V(\bar S)/\partial \bar S^n$, the prime denotes the derivative with respect to the conformal time, the bar denotes the background value, the subscript $_{,\bar S}$ is the shorthand notation of $\partial/\partial {\bar S}$, and we have omitted the arguments of $\xi(\bar S)$, $\chi(\bar S)$, and $V(\bar S)$.
From the first equation, we can obtain the following relation:
\begin{align}
        \bar \theta' \propto (a^2 \chi \bar S^2)^{-1},
        \label{eq:s_ch_cons}
\end{align}
where this corresponds to the conservation of the angular momentum, i.e. the U(1) charge. 
In the following, we assume fast rotations with $a^2 V^\fo/(\chi \bar S^2)_{,\bar S} \gg \mathcal H^2$ \footnote{In the two-field model, which we focus on in the main text, we find $V^\fo /(\chi \bar S^2)_{,\bar S} \simeq V^\fo/\bar S \simeq m_S^2$ in $\bar S \gg f_a$.
In the observable regions in Fig.~\ref{fig:ms_fa}, we find $a m_S/\mathcal H|_{\eta_\eqf} \gtrsim \mathcal O(10^{5})$.}
and a heavy radial field with $a^2 V^\fo/(\bar S \xi) \gg \mathcal H^2$ in Eq.~(\ref{eq:rbar_eom_g}).
In this case, the background solution of the radial field can be divided into two parts $\bar{S} = \bar{S}^{\text{(fast)}} + \bar{S}^{\text{(slow)}}$: the fast oscillating part $\bar{S}^{\text{(fast)}}$ with $(\ln \bar S^{\text{(fast)}})' \sim \mathcal O(\sqrt{a^2 V^\fo /(\xi \bar S)})$ and the (relatively) slowly evolving part $\bar{S}^{\text{(slow)}}$ with $(\ln \bar S^{\text{(slow)}})' \ll \mathcal O(\sqrt{a^2 V^\fo /(\xi \bar S)})$.
The fast oscillating part makes the orbit elliptical and we are interested in the case where it is negligible because of the dissipation through the interaction with the thermal bath (see Refs.~\cite{Co:2021lkc, Kawamura:2021xpu} and references therein for the dissipation process). 
For this reason, we hereafter focus on the slowly evolving part $\bar{S}=\bar{S}^{\text{(slow)}}$, which we call the circular mode.
We can obtain the following circular-mode relation by equating the effective mass term in Eq.~(\ref{eq:rbar_eom_g}) as 
\begin{align}
        &a^2 V^\fo - \frac{1}{2}\left(\chi\bar{S}^2\right)_{,\bar S} \left(\bar{\theta }'\right)^2 \simeq 0. 
        \label{eq:s_back_con}
\end{align}
This circular-mode relation is self-consistent if the other terms in Eq.~(\ref{eq:rbar_eom_g}) are negligible.
We here derive the condition for Eq.~(\ref{eq:s_back_con}) to be self-consistent.
To this end, we first estimate the timescale of $(\ln S)'$. 
Taking the time derivative of Eq.~(\ref{eq:s_back_con}) and using Eq.~(\ref{eq:thbar_eom_g}), we can obtain 
\begin{align}
    \bar S' \simeq -\frac{\mathcal H}{\mathcal Y},
    \label{eq:s_dash}
\end{align}
where 
\begin{align}
    \mathcal Y \equiv \frac{1}{6} \left[\ln \left( V^\fo \frac{(\chi \bar S^2)^2}{(\chi \bar S^2)_{,\bar S}} \right) \right]_{, \bar S}.
\end{align}
From this, we can see that Eq.~(\ref{eq:s_back_con}) is valid when 
\begin{align}
    \frac{a^2 V^\fo}{\xi} \gg \mathcal \max\left[ \left|\left(\frac{\mathcal H}{\mathcal Y}\right)'\right|, \, \left|\frac{2\mathcal H^2}{\mathcal Y} \right|, \, \left| \left(\frac{\mathcal H}{\mathcal Y}\right)^2 \frac{\xi_{,\bar S}}{\xi} \right| \right].
\end{align}
We have checked that this condition is satisfied in our fiducial setup in the main text mainly because of the fast rotations, that is, the large hierarchy between $a m_S$ and $\mathcal H$.
The physical meaning of this is that the heavy radial mode cannot be excited once dissipated and therefore it can be safely integrated out.

At the linear perturbation level, the equations of motion are
\begin{align}
&  \delta \theta ''+\left(\ln\left(a^2 \chi \bar S^2 \right)\right)' \delta \theta ' + k^2 \delta \theta -4 \bar{\theta }'\Phi' + 2 \frac{\chi_{,\bar S}}{\chi} \bar S' \bar \theta' \Phi  
- 2 \frac{\bar \theta'}{\bar S} \left[ \left(\frac{\bar{S}' }{\bar{S}} + \frac{\left(\chi_{,\bar S}\right)^2 \bar S \bar S'}{2 \chi^2} - \frac{\chi_{,\bar S \bar S} \bar S \bar S'}{2\chi} \right)\delta S  - \left( 1 + \frac{\chi_{,\bar S} \bar S}{2\chi} \right)\delta S' \right] = 0, \\
& \delta S'' +2 \mathcal{H} \delta S' + \frac{1}{\xi} \left[ \left(\xi k^2 + a^2 V^\so - \frac{1}{2} \left( \chi \bar S^2\right)_{,\bar S \bar S}\left(\bar{\theta }'\right)^2 \right) \delta S - \left(\chi \bar S^2 \right)_{,\bar S} \bar{\theta }'\delta \theta' \right] -4 \bar{S}'\Phi ' +\frac{2 a^2 V^\fo}{\xi}\Phi \nonumber \\
&\qquad\qquad\qquad\qquad\qquad\qquad\qquad\qquad
  -\frac{a^2 V^\fo \xi_{,\bar S} }{\xi^2}\delta S + \frac{2 \bar{S}' \xi_{,\bar S} }{\xi}\delta S' 
+\frac{\left(\chi \bar S^2\right)_{, \bar S}  \left(\bar{\theta }'\right)^2 \xi_{, \bar S}}{2\xi^2}\delta S
+\left(\bar{S}'\right)^2 \left(\frac{\xi_{, \bar S}}{\xi}\right)_{, \bar S}\delta S =0, 
\end{align}
where we have taken the conformal Newtonian gauge (Eq.~(\ref{eq:newton_gauge})) and used the perfect fluid condition $\Phi = \Psi$.
Substituting the background circular-mode relation Eq.~(\ref{eq:rbar_eom_g}) into the second equation, we obtain
\begin{align}
  & \delta S'' +2 \mathcal{H} \delta S' + \frac{1}{\xi} \left[ \left(\xi k^2 + a^2 V^\so - \frac{1}{2} \left( \chi \bar S^2\right)_{,\bar S \bar S}\left(\bar{\theta }'\right)^2 \right) \delta S - \left(\chi \bar S^2 \right)_{,\bar S} (\bar{\theta }'\delta \theta' - (\bar \theta')^2 \Phi) \right] \nonumber \\
   & \qquad\qquad\qquad\qquad\qquad\qquad\qquad\qquad\qquad\qquad
   -4 \bar{S}'\Phi' + \frac{2 \bar{S}' \xi_{,\bar S} }{\xi}\delta S' 
+\left(\bar{S}'\right)^2 \left(\frac{\xi_{, \bar S}}{\xi}\right)_{, \bar S}\delta S \simeq 0.
\label{eq:dS_eom}    
\end{align}
Similar to the background, we are interested in the circular mode for the perturbation, which does not break the circular orbit of the rotational field motion.
Neglecting terms that include $\bar{S}'$, $\delta S'$, or $\delta S''$, we obtain the following relation between $\delta S$, $\delta \theta'$, and $\Phi$ for the circular mode:
\begin{align}
        &\left(\xi k^2 + a^2 V^\so - \frac{1}{2}\left(\chi \bar S^2 \right)_{,\bar S\bar S}\left(\bar{\theta }'\right)^2 \right)\delta S 
        \simeq \left(\chi \bar S^2 \right)_{,\bar S} \left(\bar{\theta }'\delta \theta' - \left(\bar{\theta}'\right)^2 \Phi \right),
        \label{eq:dS_cond}
\end{align}
where we have kept the $k^2$ term by assuming $k \gg \mathcal H$.
We have checked that this relation is self-consistent in our setup in the main text because of the large hierarchy $am_S/\mathcal H \gg 1$, similar to the background case.

We now see how $w$ and $c_s$ are expressed in the case where only the circular mode exists.
For the Lagrangian of Eq.~(\ref{eq:L_gen_S}), the energy momentum tensor is given by~\cite{Weinberg:2008zzc} 
\begin{align}
        T^\mu_{\ \nu} = \delta^\mu_{\ \nu} \left[ -\frac{1}{2} (\xi(S) \partial^\sigma S \partial_\sigma S + \chi(S) S^2 \partial^\sigma \theta \partial_\sigma \theta) - V(S) \right] + \xi(S) \partial^\mu S \partial_\nu S + \chi(S) S^2 \partial^\mu \theta \partial_\nu \theta.
\end{align}
From this expression, the background energy density and pressure are given by
\begin{align}
        -\bar T^0_{\ 0} &= \bar \rho = \frac{1}{2a^2} \left[\xi(\bar S) (\bar S')^2 + \chi(\bar S)\bar S^2 (\theta')^2 \right]  + V(\bar S), \\
        \frac{1}{3}\bar T^i_{\ i} &= \bar p= \frac{1}{2a^2} \left[\xi(\bar S) (\bar S')^2 + \chi(\bar S)\bar S^2 (\theta')^2 \right]   - V(\bar S).
\end{align}
Substituting Eq.~(\ref{eq:s_back_con}) into this, we can express $w$ as 
\begin{align}
        \label{eq:w}
        w = \frac{\bar p}{\bar \rho} \simeq \frac{V^\fo -  (\chi \bar S^2)_{,\bar S}V/(\chi \bar S^2)}{V^\fo + (\chi \bar S^2)_{,\bar S}V/(\chi \bar S^2)}.
\end{align}
where we have assumed $a^2(\chi \bar S^2) V^\fo \mathcal Y^2/((\chi \bar S^2)_{,\bar S} \xi \mathcal H^2) \gg 1$, which is the case in our setup in the main text due to $am_S/\mathcal H \gg 1$.
Next, let us see the perturbations of the energy density and the pressure:
\begin{align}
        \label{eq:rho_exa}
        -\delta T^0_{\ 0} &= \delta \rho = V^\fo \delta S+ \frac{1}{a^2} \left(  \frac{\xi_{,\bar S} \left(\bar{S}'\right)^2+(\chi\bar S^2)_{,\bar S} (\bar \theta')^2}{2}\delta S   + \xi \bar{S}' \left(\delta S'- \bar{S}'\Phi\right)+ \chi \bar{S}^2 \left(\bar{\theta }'\delta \theta ' - (\bar \theta')^2\Phi \right)\right),\\ 
        \label{eq:P_exa}       
        \frac{1}{3} \delta \bar T^i_{\ i} &= \delta p = -V^\fo \delta S+ \frac{1}{a^2} \left(  \frac{\xi_{,\bar S} \left(\bar{S}'\right)^2+(\chi\bar S^2)_{,\bar S} (\bar \theta')^2}{2}\delta S   + \xi \bar{S}' \left(\delta S'- \bar{S}'\Phi\right)+ \chi \bar{S}^2 \left(\bar{\theta }'\delta \theta ' - (\bar \theta')^2\Phi \right)\right).
\end{align}
Using Eqs.~(\ref{eq:s_back_con}) and (\ref{eq:dS_cond}), we can obtain the perturbations for the circular mode,
\begin{align}
        \label{eq:d_rho}
        \delta \rho &\simeq 2V^\fo \delta S + \frac{\chi \bar S^2}{(\chi \bar S^2)_{,\bar S}} \left( \frac{\xi k^2}{a^2} + V^\so - V^\fo \frac{\left(\chi \bar S^2 \right)_{,\bar S\bar S}}{\left(\chi \bar S^2 \right)_{,\bar S}}\right)\delta S, \\
        \label{eq:d_p} 
        \delta p & \simeq \frac{\chi \bar S^2}{(\chi \bar S^2)_{,\bar S}} \left( \frac{\xi k^2}{a^2} + V^\so - V^\fo \frac{\left(\chi \bar S^2 \right)_{,\bar S\bar S}}{\left(\chi \bar S^2 \right)_{,\bar S}}\right)\delta S,
\end{align}
where we have neglected the subdominant terms due to the large hierarchy $am_S/\mathcal H \gg 1$ in our setup in the main text.
Then, we finally obtain the sound speed, 
\begin{align}
        \label{eq:cs2}
        c_s^2 \equiv \frac{\delta p}{\delta \rho} \simeq \frac{ V^\so - V^\fo \frac{\left(\chi \bar S^2 \right)_{,\bar S\bar S}}{\left(\chi \bar S^2 \right)_{,\bar S}}}{ V^\so +  V^\fo \left(\frac{2 (\chi\bar S^2)_{,\bar S}}{\chi \bar S^2}-\frac{\left(\chi \bar S^2 \right)_{,\bar S\bar S}}{\left(\chi \bar S^2 \right)_{,\bar S}}\right)},
\end{align}
where we have further neglected $\mathcal O(\xi k^2/(a^2 \tilde m^2))$ correction in the RHS with $\tilde m^2 \equiv \max[2 V^\fo(\chi \bar S^2)_{,\bar S}/(\chi \bar S^2) ,V^\so- V^\fo \left(\chi \bar S^2 \right)_{,\bar S\bar S}/\left(\chi \bar S^2 \right)_{,\bar S})]$.
By combining Eqs.~(\ref{eq:s_dash}), (\ref{eq:w}), and (\ref{eq:cs2}), we can easily confirm that the adiabatic relation $c_s^2 = w-w'/(3\mathcal H (1+w))$ is satisfied in the case that we focus on.

\section{Two-field model}
\label{smsec:two_field_model}

In this section, we discuss the supersymmetric UV completion of the effective Lagrangian in Eq.~(\ref{eq:Leff}) and the field evolution in the effective Lagrangian.
The model has two $\mathrm{U}(1)$-charged fields $P$ and $\bar{P}$ and a $\mathrm{U}(1)$ neutral field $X$. The K\"ahler potential and the superpotential are given by 
\begin{align}
    K =& |P|^2 + |\bar{P}|^2 + |X|^2, \\
    W =& \lambda X (P \bar{P} - v^2),
\end{align}
where $\lambda$ is a dimensionless coupling constant and $v$ is a mass-dimensional parameter controlling the vacuum expectation value of $P \bar{P}$.  
The supersymmetric potential of $P$, $\bar{P}$, and $X$ is given by 
\begin{align}
    V = \lambda^2|P\bar{P}-v^2|^2 + \lambda^2|X|^2 (|P|^2 + |\bar{P}|^2).
\end{align}
The term $\lambda^2|P\bar{P}-v^2|^2$ fixes $P$ and $\bar{P}$ to a flat direction $P \bar{P} = v^2$.
The flat direction is lifted by soft supersymmetry breaking masses,
\begin{align}
V_{\rm soft} = m_P^2 |P|^2 + m_{\bar{P}}^2|\bar{P}|^2.
\end{align}
We assume a hierarchy $\lambda v \gg m_P, \, m_{\bar{P}}$. Then $X$ and a linear combination of $P$ and $\bar{P}$ (expanded around an expectation value along the flat direction) obtain a large mass about $\frac{\lambda}{2}{\rm max}(P,\bar{P}) (> \lambda v \gg m_{P}, \, m_{\bar{P}})$. We may integrate out the heavy linear combination via $\bar{P} = v^2/P$.
Then, the kinetic term becomes
\begin{align}
    - \partial^\mu P^\dag \partial_\mu P - \partial^\mu \bar{P}^\dag \partial_\mu \bar{P} \rightarrow - \left(1 + \frac{v^4}{|P|^4} \right) \partial^\mu P^\dag \partial_\mu P , 
\end{align}
and the potential becomes
\begin{align}
    m_P^2|P|^2 +  m_{\bar{P}}^2 |\bar{P}|^2 \rightarrow  \left( 1 + \frac{m_{\bar{P}}^2}{m_P^2} \frac{v^4}{|P|^4} \right) m_P^2|P|^2 .
\end{align}
After adding a constant term so that the potential energy vanishes at the vacuum,  the effective Lagrangian is then given by Eq.~(\ref{eq:Leff}) in the main text, with $1+d = m_{\bar{P}}/m_P$, $m_S= m_P$, and $v = f_a/2$. When the theory enjoys approximate $Z_2$ symmetry $P\leftrightarrow \bar{P}$, $d \ll 1$ and the sudden change of $c_s$ is realized.

In the following, we discuss the evolution of the field in the two-field model.
For convenience, we define $r$ as 
\begin{align}
    2|P|^2\left( 1 + \frac{v^4}{|P|^4} \right) = r^2,
    \label{eq:p_r_rel}
\end{align}
and rewrite the Lagrangian Eq.~(\ref{eq:Leff}) in the main text as 
\begin{align}
    \mathcal L = - \frac{\xi(r)}{2} \partial^\mu r \partial_\mu r  - \frac{r^2}{2}\partial^\mu \theta \partial_\mu \theta - V(r),
    \label{eq:Leff_r}
\end{align}
where $r>2v$ and 
\begin{align}
    \xi(r) &= \frac{r^4}{r^4 - 16 v^4},\\
    V(r) &= \frac{1}{4}m_S^2 \left( \sqrt{r^2 + \sqrt{r^4 - 16 v^4}} - \frac{4 (1+d) v^2}{\sqrt{r^2 + \sqrt{r^4 - 16 v^4}}}\right)^2.
    \label{eq:pot_in_r}
\end{align}
Note that, for $d = 0$, the potential becomes $V(r) = m_S^2 (r^2-4v^2)/2$.
The advantage of this field redefinition is that we can substitute $\chi = 1$ into the expressions in the previous section, which significantly simplifies them.
Similar to the previous section, we focus on the mode that does not break the circular orbit of the rotating field in the following. 
Then, from Eq.~(\ref{eq:s_back_con}), $\theta'$ and $\bar r$ follow
\begin{align}
    (\bar{\theta}')^2 = a^2 V^\fo/\bar r.
    \label{eq:th_v}
\end{align}
Also, from the charge conservation Eq.~(\ref{eq:s_ch_cons}), we obtain 
\begin{align}
   \bar r^2 {\bar \theta}' \propto a^{-2}.
    \label{eq:ch_cons}
\end{align} 
From Eqs.~(\ref{eq:w}) and (\ref{eq:cs2}), we can express $w$ and $c_s^2$ as 
\begin{align}
  \label{eq:w_sm}
     w &= \frac{V^\fo - 2V/r}{V^\fo + 2V/r}, \\
  c_s^2 &= \frac{V^\so - V^\fo/r }{ V^\so + 3 V^\fo /r }.
  \label{eq:cs}
\end{align}
Note that we can also derive this expression of $c_s^2$ by using the adiabatic relation $c_s^2 = w - w'/(3 \mathcal H(1+w))$ with Eq.~(\ref{eq:w_sm}).

\begin{figure}%[t] 
        \centering \includegraphics[width=0.45\columnwidth]{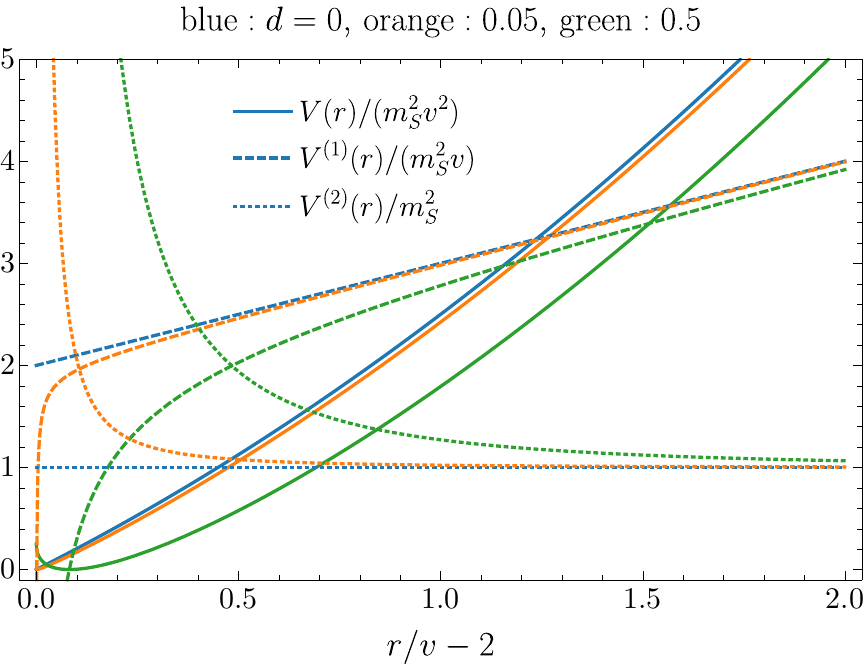}
        \caption{ 
        The potential given by Eq.~(\ref{eq:pot_in_r}) and its derivatives. 
        }
        \label{fig:pot}
\end{figure}

Figure~\ref{fig:pot} shows the potential and its derivatives.
From this figure, we can see that, for a smaller $d$, the deviations of the potential derivatives from the ones of the quadratic potential appear at smaller $r$.
This indicates that, for smaller $d$, the moving distance of $r$ during the transition from the MD to the KD era becomes smaller and therefore the transition becomes more sudden.

One might wonder why the sudden transition occurs even though the potential in the original Lagrangian, Eq.~(\ref{eq:Leff}) in the main text, does not have a sharp feature. 
We here address this question.
The behavior of this model without the field redefinition to $r$ is discussed in Ref.~\cite{Co:2021lkc}. 
To match the convention in the previous section and Ref.~\cite{Co:2021lkc}, we define $S \equiv \sqrt{2}|P|$.
Then, we can rewrite the two-field model Lagrangian, Eq.~(\ref{eq:Leff}) in the main text, as
\begin{align}
    {\cal L} &=  -\frac{\chi(S)}{2}(\partial^\mu S \partial_\mu S + S^2 \partial^\mu \theta \partial_\mu \theta) - \frac{m^2_S}{2} \left( S - \frac{2(1+d)v^2}{S}\right)^2,
    \label{eq:Leff_in_S}
\end{align}
where $\chi(S) = 1 + 4v^4/S^4$. 
This Lagrangian corresponds to the case with $\xi = \chi$ in Eq.~(\ref{eq:L_gen_S}).
Note that the potential minimum is at $S = \sqrt{2(1+d)}v$, which coincides with the minimum of $(\chi \bar S^2)$ when $d =0$.
From Eq.~(\ref{eq:s_back_con}), the background relation for the circular mode is given by
\begin{align}
    a^2 V^\fo &= \frac{1}{2}\left(\chi \bar S^2 \right)_{,\bar S} (\theta')^2 \nonumber \\
    &= \frac{\bar S^4 - 4v^4}{\bar S^3} (\theta')^2.
    \label{eq:dv_s_int}
\end{align}
Note that this equation can also be obtained by changing $\bar r$ in Eq.~(\ref{eq:th_v}) to $\bar S$ through Eq.~(\ref{eq:p_r_rel}).
The U(1)-charge conservation, Eq.~(\ref{eq:ch_cons}), is given by 
\begin{align}
    \label{eq:u1_cons_s}
    \chi \bar S^2\bar \theta' \propto a^{-2}.
\end{align}
As $\bar S$ approaches the potential minimum $\sqrt{2(1+d)}v$, $(\chi \bar S^2)_{,\bar S}/\bar S$ becomes small and consequently suppresses the RHS of Eq.~(\ref{eq:dv_s_int}) because $\theta'$ does not change rapidly when $\bar S$ is close to the potential minimum, which can be seen from Eq.~(\ref{eq:u1_cons_s}).
Since Eq.~(\ref{eq:dv_s_int}) physically means the balance of the centripetal force (LHS) and the centrifugal force (RHS), the suppression of the RHS corresponds to the decrease of the centrifugal force for the field $S$, which effectively accelerates the rolling down of $S$ to the potential minimum.
For a smaller $d$, the matter-to-kination transition occurs around the field value closer to the zero point of $(\chi \bar S^2)_{,\bar S}$, where the rolling down of $S$ is faster. 
Due to this accelerated rolling down, the sharp transition is realized for a small $d$.
Note that, for a large $d$ (say, $d\sim 1$), $(\chi \bar S^2)_{,\bar S}/\bar S$ does not become small even at the potential minimum, which means that the acceleration of the roll down to the potential minimum is small in that case.
On the other hand, for a small $d$, the potential minimum becomes close to the zero-point of $(\chi \bar S^2)_{,\bar S}$ and they coincide in the $d \rightarrow 0$ limit. 
This closeness is required for the sharp transition from the MD to the KD era.
The origin of this sharp evolution feature in $S$ comes from the noncanonical form of the axion kinetic term, $(\chi S^2) \partial^\mu \theta \partial_\mu \theta$. 
The field redefinition of $S$ to $r$ changes $(\chi S^2) \partial^\mu \theta \partial_\mu \theta$ to $r^2 \partial^\mu \theta \partial_\mu \theta$, and then the sharp-evolution feature in the field value of $S$ is moved into the sharp feature of the potential in $r$.

\section{Evolution of scale factor, Hubble parameter, and gravitational potential}
\label{smsec:a_and_h}

In this section, we summarize the evolution of the scale factor, the Hubble parameter, and the gravitational potential through the transitions of multiple universe eras with different $w$.

\subsection{Scale factor and Hubble parameter}

First, we discuss the evolution of the scale factor and the Hubble parameter.
Although we focus on the kination era ($w=1$) after the MD era in the main text, we here summarize the equations in the situation that includes a general era with $w > 1/3$ ($w$D era). 
Then, we focus on the transitions of the first RD era $\rightarrow$ MD era $\rightarrow$ $w$D era $\rightarrow$ the second RD era.
We consider the two-component fluid with radiation and the matter whose equation-of-state parameter instantaneously changes from zero to $w (> 1/3)$ at $\eta_w$.
We define $\eta_\eqf (< \eta_{w})$ as the first equality time for the radiation and the matter at the beginning of the MD era and $\eta_\eqs (> \eta_{w})$ as the second equality time at the end of the $w$D era.
As in the main text, we assume the effective degree of freedom is constant during the transitions for simplicity.
Using the continuities of the scale factor and the Hubble parameter at $\eta_w$, we obtain the following expressions in $\eta \ll \eta_\eqs$:
\begin{align}
        \label{eq:a_o_aeq}
        \frac{a(\eta)}{a(\eta_\eqf)} &= 
        \begin{cases}
                \cfrac{\eta (\eta + 2 \eta_\stf)}{\eta_\stf^2}  & (\eta < \eta_{w}) \\
                \left( \cfrac{\eta- \eta_\diamond}{\eta_{w} - \eta_\diamond} \right)^{2/(1+3w)} \cfrac{\eta_{w}( \eta_{w} + 2 \eta_\stf)}{\eta_\stf^2}   & (\eta_w \leq \eta \ll \eta_\eqs)
        \end{cases},
        \\
        \mathcal H &= 
        \begin{cases}
                \cfrac{2(\eta + \eta_\stf)}{\eta (\eta + 2 \eta_\stf)}  & (\eta < \eta_{w}) \\
                \cfrac{2}{(1+3w)(\eta - \eta_\diamond)} & ((\eta_w \leq \eta \ll \eta_\eqs))
        \end{cases},
        \\
        \eta_\diamond &= \eta_{w} - \frac{\eta_{w}(\eta_{w} + 2 \eta_\stf)}{(1+3w) (\eta_{w} + \eta_\stf)},
\end{align}
where $\eta_\stf = \eta_\eqf/(\sqrt{2}-1)$.
Actually, these expressions are also applicable for 
 $w=1/3$ and reproduce the expressions in Ref.~\cite{Inomata:2020lmk}.

Next, we discuss the evolution after the instantaneous transition.
In general, in the case of $w > 1/3$, the radiation dominates the universe again after a while. 
To be concrete, we here set $w = 1$, which corresponds to the case in the main text. 
Note that $\eta_{w=1} = \eta_\kk$ in the main text.
Using the fact that the energy densities of radiation and the rotating (kination) field are $\rho_\rr \propto 1/a^4$ and $\rho_\kk \propto 1/a^6$, we can express the Friedmann  equation as 
\begin{align}
        \mathcal H^2 = \frac{\mathcal H^2_\kk}{1+R} \left( \left(\frac{a_\kk}{a(\eta)}\right)^2 + R %\frac{a_\kk}{a_\eqf} 
         \left(\frac{a_\kk}{a(\eta)}\right)^4 \right) \ \qquad (\eta > \eta_\kk),
\end{align}
where $a_{\bullet}$ and $\mathcal H_{\bullet}$ denote $a(\eta_\bullet)$ and $\mathcal H(\eta_\bullet)$ with $\bullet$ being an arbitrary subscript, $R \equiv a_\kk/a_\eqf (= \eta_\kk(\eta_\kk + 2 \eta_\stf)/\eta_\stf^2)$, and we have used $\rho_\rr(\eta_\kk)/\rho_\kk(\eta_\kk) = a_\eqf/a_\kk$.
Solving this equation, we obtain 
\begin{align}
      \frac{a(\eta)}{a_\eqf} = \sqrt{ R^2 \frac{\mathcal H^2_\kk}{1+R} (\eta+ \eta_\triangle)^2 - R^3},
        \label{eq:a_o_a1}
\end{align}
where $\eta_\triangle$ is an integration constant. 
Then, the Hubble parameter becomes 
\begin{align}
        \mathcal H = \frac{\eta+\eta_\triangle}{(\eta+ \eta_\triangle)^2 - R(1+R)/\mathcal H^2_\kk}.
\end{align}
By imposing the continuity of $a$ and $\mathcal H$ at $\eta_\kk$, we express $\eta_\triangle$ as 
\begin{align}
        \eta_\triangle = \frac{\eta_\kk^2(\eta_\kk + 3 \eta_\stf)}{2 \eta_\stf^2}.
\end{align}
Using Eq.~(\ref{eq:a_o_a1}) and $a_\kk/a_\eqf = (a_\eqs/a_\kk)^2$, we can relate $\eta_\eqs$ to $\eta_\eqf$ and $\eta_\kk$ as
\begin{align}
        \label{eq:eqs_o_eqf}
        \frac{\eta_\eqs}{\eta_\eqf} \simeq \frac{1}{2(7+5\sqrt{2})} \left( \frac{\eta_\kk}{\eta_\eqf}\right)^3,
\end{align}
where we have assumed $\eta_\kk \gg \eta_\eqf$.

\subsection{Gravitational potential}

Next, we discuss the evolution of the gravitational potential.
Similar to the previous subsection, we consider the $w$D era after the MD era for generality.
To avoid heavy computational costs in the calculation of the induced GWs, it is important to derive the analytical expression of the gravitational potential especially after the $w$D era begins, where the poltergeist mechanism occurs.
To this end, we here consider the case of $w = c_s^2$, which corresponds to the situation after the transition from the MD era to the $w$D era has been completed.
In this case, the equation of motion for $\Phi$ (Eq.~(\ref{eq:phi_eom}) in the main text) becomes 
\begin{align}
        \Phi''_{\bm k} + 3(1+w)\mathcal H \Phi'_{\bm k} + w k^2 \Phi_{\bm k} = 0.
        \label{eq:phi_eom_w}
\end{align}
We assume the same \emph{background} evolution as in the previous subsection, where the instantaneous transition from the MD era to the $w$D era occurs at $\eta_w$ with $\eta_w \gg \eta_\eqf$.
Then, the solution of Eq.~(\ref{eq:phi_eom_w}) becomes
\begin{align}
        \Phi_{\bm k}(\eta) = C_{1\bm k} \tilde J (w; k \tilde \eta) + C_{2 \bm k} \tilde Y(w; k \tilde \eta),
\end{align}
where $\tilde \eta \equiv  \eta- \eta_\diamond$ and 
\begin{align}
        \tilde J(w; k\tilde \eta) &\equiv (k \tilde \eta)^{-\alpha} J_\alpha (\sqrt{w} k \tilde \eta), \\ 
        \tilde Y(w; k \tilde \eta) &\equiv (k\tilde \eta)^{-\alpha} Y_\alpha (\sqrt{w} k \tilde \eta), \\  
        \alpha &\equiv \frac{1}{2} \left( \frac{5+3w}{1+3w} \right). 
\end{align}
The $J_\alpha$ and $Y_\alpha$ are the Bessel functions of the first and the second kind, respectively.
The coefficients $C_{1,2 \bm k}$ are determined with the boundary conditions of $\Phi$ and $\Phi'$ at $\eta_w$ as 
\begin{align}
        \label{eq:c12}        
        C_{1 \bm k}&=  \frac{\Phi_{\bm k}(\eta_{w}) \tilde Y'({w};y_{w})- \Phi'_{\bm k}(\eta_{w}) \tilde Y({w};y_{w}) }{\tilde J({w};y_{w}) \tilde Y'({w};y_{w}) - \tilde J'({w};y_{w}) \tilde Y({w};y_{w})},\\
        C_{2 \bm k}&= \frac{\Phi_{\bm k}(\eta_{w}) \tilde J'({w};y_{w})- \Phi'_{\bm k}(\eta_{w})\tilde J({w};y_{w})}{\tilde Y({w};y_{w}) \tilde J'({w};y_{w}) - \tilde Y'({w};y_{w}) \tilde J({w};y_{w})},
\end{align}
where $y_{w} = k \tilde \eta_{w} = k(\eta_{w}-\eta_\diamond)$, $\tilde J'({w};y) = \partial \tilde J({w};k \tilde \eta)/\partial \tilde \eta$, and the same for $\tilde Y'({w};y)$.
We here keep $\Phi'(\eta_w)$, which is zero in the instantaneous transition limit, because we use this formula to take into account the suppression of $\Phi$ during the transition.
We discuss this point in the next section.

In the superhorizon limit during the MD era, the gravitational potential is related to the curvature perturbation as $\Phi_{\bm k} = -(3/5)\zeta_{\bm k}$.
Taking into account this, we define the transfer function as
\begin{align}
        \Phi_{\bm k} = - \frac{3}{5}\zeta_{\bm k} T(k \eta).
        \label{eq:phi_trans}
\end{align}
The concrete expression of the transfer function is given by
\begin{align}
        \label{eq:trans}
        T(x) &= \left\{
        \begin{array}{ll}
               M(x_\eqf)  &  (\eta_\eqf \ll \eta < \eta_{w}) \\
               M(x_\eqf)( D_1({w}, y_{w}) \tilde J ({w}; y) + D_{2}({w}, y_{w}) \tilde Y({w}; y)) & (\eta \geq \eta_{w})
        \end{array},
        \right.
\end{align}
where $D_{1,2}({w},y_{w}) \equiv C_{1,2 {\bm k}}/(-3\zeta_{\bm k}/5)$, $x_\eqf \equiv k \eta_\eqf$, and $M(x)$ is the suppression factor for the perturbations that enter the horizon during the first RD era~\cite{Bardeen:1985tr,Dodelson:1282338}:
\begin{align}        
        M(x_\eqf) & =  \frac{\ln \left( 1+0.146\, x_\eqf \right)}{\left(0.146\, x_\eqf \right)}\left[ 1 + 0.242\, x_\eqf + \left(1.01\, x_\eqf \right)^2 
+ \left(0.341\, x_\eqf \right)^3 + \left(0.418\, x_\eqf \right)^4 \right]^{-0.25}. \label{eq:m_eq}
\end{align}
Note that $M(x_\eqf) \simeq 1$ for $x_\eqf \ll 1$ and $\Phi' = 0$ in $\eta_\eqf \ll \eta < \eta_{w}$.
In particular, for the scales that enter the horizon during the RD era before the MD era, $M(x_\eqf)$ takes into account the suppression of the gravitational potentials during the oscillation in the RD era and its motion
after entering the MD era. The oscillation of the gravitational potential stops shortly after entering the MD era and the initial condition $\Phi'=0$ is a good approximation as long as the initial time is taken to be not close to the RD-to-MD transition.

\section{Calculation of induced gravitational waves}
\label{smsec:induced_gw}

In this section, we explain how to calculate the GWs induced by the poltergeist mechanism numerically and analytically in our setup.

\subsection{Basic formulas for induced gravitational waves}

First, we summarize the basic formulas for the induced GWs in the transitions of the first RD era $\rightarrow$ MD era $\rightarrow$ $w$D era $\rightarrow$ the second RD era with the instantaneous transition between the MD era and the $w$D era.
See e.g. Refs.~\cite{Kohri:2018awv,Domenech:2019quo} for the intermediate steps in the derivation of the formulas.

We expand the tensor perturbations as
\begin{align}
	h_{ij}(\bm x) =  \sum_{\lambda = +,\times}\int \frac{\dd k^3}{(2\pi)^3} e^\lambda_{ij}(\hat{\bm{k}})\, h^\lambda_{\bm k} \, \ee^{i \bm{k} \cdot \bm{x}},
\end{align}
where the polarization tensor satisfies $k^i e^\lambda_{ij} = 0$, $\delta^{ij} e^\lambda_{ij}=0$, and $\delta^{ik} \delta^{jl}e^\lambda_{kl} e^{\lambda'}_{ij}=\delta^{\lambda \lambda'}$.
With these tensor Fourier modes, we define the tensor power spectrum as 
\begin{align}
	\vev{h^\lambda_{\bm k} h^{\lambda'}_{\bm k'}} = (2\pi)^3 \delta(\bm k + \bm k') \delta^{\lambda \lambda'} \frac{2\pi^2}{k^3} \mathcal P_h(\eta, k).
\end{align}
From the Einstein equation, we can obtain the equation of motion for tensor perturbations, given by Eqs.~(\ref{eq:h_eom}) and (\ref{eq:s}) in the main text.
To solve the equation of motion, we here introduce the Green's function that satisfies $\hat {\mathcal N} G_k(\eta,\bar \eta) = \delta (\eta - \bar \eta)$, where 
\begin{align}
        \hat{\mathcal N} \equiv \frac{\partial^2}{\partial \eta^2} + 2 \mathcal H \frac{\partial}{\partial \eta} + k^2.
\end{align}
After some calculation with the Green's function and the transfer function defined by Eq.~(\ref{eq:phi_trans}), we obtain the induced tensor power spectrum as 
\begin{align}
        \mathcal P_h(\eta, k) = 4 \int^\infty_0 \dd v \int^{1+v}_{|1-v|} \dd u \left( \frac{4v^2 - (1+v^2-u^2)^2}{4 u v} \right)^2 I^2(u,v,x) \mathcal P_\zeta(k u) \mathcal P_\zeta(k v),
        \label{eq:p_h_uv}
\end{align}
where $x = k \eta$ and 
\begin{align}
  I(u,v,x) = \int^x_0 \dd \bar x\, k G_k(\eta, \bar \eta) f(u,v,\bar x).
  \label{eq:I_def}
\end{align}
The source function $f$ is given by 
\begin{align}
        f(u,v,\bar x) = \frac{3}{25(1+w)} \left[ 2(5+3w) T(u\bar{x})T(v\bar{x}) +4 \mathcal H^{-1} \left(T'(u\bar{x})T(v\bar{x}) + T(u\bar{x})T'(v\bar{x})\right) + 4 \mathcal H^{-2} T'(u\bar{x})T'(v\bar{x}) \right],
\label{eq:f_def}
\end{align}
where $T'(x) = \partial T(x)/\partial \eta = k \partial T(x)/\partial x$.
For later convenience, we here show another expression of Eq.~(\ref{eq:p_h_uv}) with $t = u+v-1$ and $s=u-v$:
\begin{align}
    \mathcal P_h(\eta, k) = 2 \int^\infty_0 \dd t \int^{1}_{-1} \dd s \left[ \frac{t(2+t)(s^2-1)}{(1-s+t)(1+s+t)} \right]^2 I^2(u,v,x) \mathcal P_\zeta(k u) \mathcal P_\zeta(k v).
    \label{eq:p_h_st}
\end{align}
An advantage of this expression is that the integration region is simple. 
In particular, for the scale of $k \ll k_*$, the dominant contribution of the poltergeist mechanism comes from the squeezed limit, which corresponds to $t \gg 1$.  

The integration region of Eq.~\eqref{eq:I_def} can be divided into four corresponding to the first RD, MD, $w$D, and the second RD eras.  In the following, we explain that the main contribution to the induced GWs comes from the $w$D era when the MD$\to w$D transition is sudden enough. 
\begin{itemize}
\item During the first RD era, the source term is not enhanced because the density perturbations remain small due to the pressure of the fluid.
\item During the MD era, the source term becomes larger than that during the first RD era because of the growth of the density perturbations on subhorizon scales. 
However, the source scalar perturbations do not oscillate. 
Due to this, the source term during the MD era is still smaller than that soon after the MD era, which is further boosted by the fast oscillations of the density perturbations and realizes the poltergeist mechanism.
In addition, the tensor perturbations sourced by the non-oscillating scalar perturbations are not the same as the propagating GWs. They are standing waves during the MD era. 
It can be converted to the propagating GWs after the MD era, and its conversion rate sensitively depends on the transition timescale compared to the frequencies of the sourced tensor perturbations~\cite{Inomata:2019zqy}. In our case, the conversion is efficient as we are interested in the sudden transition, but the strength of this contribution to the GWs just after the MD era is upper bounded by that calculated in Ref.~\cite{Assadullahi:2009nf} (see Ref.~\cite{Kohri:2018awv} for the revised numerical coefficient). It is subdominant compared with the GWs induced during the $w$D era.  
For these reasons, the contribution from the MD era is negligible compared to that from the poltergeist mechanism, which occurs after the MD era. 

\item During the second RD era, the gravitational potential is determined by the radiation perturbations that experience the first RD$\rightarrow$MD$\rightarrow w$D$\rightarrow$ the second RD. Remember that the radiation in the second RD era is not produced by the decay of the axion field, but they are just the remnant of the first RD era. Therefore, it does not inherit the growth of the density perturbations during the MD era.  While the axion rotation perturbations grow during the MD era, the radiation perturbations do not due to their pressure. 
Since the poltergeist mechanism partly relies on the growth of the density perturbations during the MD era, the GW production during the second RD era is negligible compared to the poltergeist mechanism during the $w$D era at least for the wavenumber range of our interest, $k\eta_{\text{eq},2}\gtrsim 1$. 
\end{itemize}

For these reasons, we focus on the integration kernel in the $w$D era and in particular the contribution from the poltergeist mechanism. 
Specifically, we can approximate $I$ as 
\begin{align}
         I(u,v,x) &\simeq \int^x_{x_w} \dd \bar x\, k G_k(\eta, \bar \eta) f(u,v,\bar x, x_w) \nonumber \\
         & = \frac{\pi}{2} y^{-\beta} \left[ Y_\beta(y) \mathcal I_{J}(u,v,x) -  J_\beta(y) \mathcal I_{Y}(u,v,x) \right],
         \label{eq:i_kernel}
\end{align}
where $x_w= k\eta_w$, $y = k\tilde \eta$, $G_k$ is the Green's function during the $w$D era ($\eta > \eta_w$):
\begin{align}
        \label{eq:green}
         k G_k(\eta,\bar \eta) &= \frac{\pi}{2} \frac{\bar y^{1+\beta}}{y^\beta} (J_\beta(\bar y) Y_\beta(y) - J_\beta(y) Y_\beta(\bar y)), \\ 
      \beta &\equiv  \frac{3(1-w)}{2(1+3w)},
\end{align}
and $\cal I_J$ and $\cal I_Y$ are defined as 
\begin{align}
       \label{eq:calI_J}
       \mathcal I_{J}(u,v,x) &\equiv  \int^x_{x_w} \dd \bar x \, \bar y^{\beta + 1} J_\beta(\bar y) f(u,v,\bar x), \\ 
       \mathcal I_{Y}(u,v,x) &\equiv  \int^x_{x_w} \dd \bar x \, \bar y^{\beta + 1} Y_\beta(\bar y) f(u,v,\bar x).
       \label{eq:calI_Y}
\end{align}
When the induced tensor perturbation is deeply inside the horizon during the $w$D era ($1/k \ll \eta < \eta_\eqs$), we can approximate $I$ as 
\begin{align}
        I(u,v,x(\gg 1))
         & \simeq -\sqrt{\frac{\pi}{2}} y^{-\beta-1/2} \left[ \sin\left( \frac{1}{4}(1 + 2 \beta)\pi - y \right) \mathcal I_{J}(u,v,x) + \cos\left( \frac{1}{4}(1 + 2 \beta)\pi - y \right) \mathcal I_{Y}(u,v,x) \right].
\end{align}
Then, the oscillation average of $I^2$ can be expressed as 
\begin{align}
       \overline{I^2(u,v,x(\gg 1))}
         & \simeq \frac{\pi}{4} y^{-2\beta-1} \left[ \mathcal I_{J}^2(u,v,x) + \mathcal I_{Y}^2(u,v,x) \right],
\end{align}
where the overline denotes the oscillation average.

The energy density of GWs (per log bin, $\ln k$) is expressed as 
\begin{align}
	\rho_\GW(\eta,k) = \frac{k^2}{8a^2}\overline{\mathcal P_h(\eta,k)},
\end{align}
and the energy density parameter of GWs is given by 
\begin{align}
        \Omega_\GW(\eta,k) &\equiv \frac{\rho_\GW(\eta,k)}{\rho_\tot(\eta)}= \frac{1}{24} \left( \frac{k}{\mathcal H}\right)^2 \overline{\mathcal P_h(\eta,k)}, 
\end{align}
where $\rho_\tot$ is the total energy density of the universe.
In the following, we derive the energy density parameter for GWs induced during the $w$D era with $w > 1/3$, followed by the second RD era.
Even during the $w$D era, the GW-radiation energy ratio, $\rho_\GW/\rho_\rr$, approaches a constant value for subhorizon modes because the GW production of the poltergeist mechanism on subhorizon scales stops after a while and the induced GWs behave as radiation on subhorizon scales.\footnote{The superhorizon tensor modes continue to be sourced by the scalar perturbations until the tensor modes enter the horizon during the $w$D era~\cite{Domenech:2020kqm}.}
The GW-radiation energy ratio during the $w$D era is given by 
\begin{align}
        \frac{\rho_\GW(\eta,k)}{\rho_\rr(\eta)} \simeq \frac{\rho_w(\eta)}{\rho_\rr(\eta)} \Omega_\GW(\eta,k) = \frac{1}{24} \left( \frac{(1+3w)y}{2} \right)^2 \frac{a_w}{a_\eqf} \left( \frac{a_w}{a(\eta)} \right)^{3(1+w)-4} \overline{\mathcal P_h(\eta,k)} \ \ (\eta_w < \eta \ll \eta_\eqs),
        \label{eq:gw_r_ratio}
\end{align}
where $\rho_w$ is the energy density of the dominant matter in the $w$D era.
We have also used $\rho_\tot \simeq \rho_w$ during the $w$D era and $\rho_w(\eta)/\rho_\rr(\eta) = (a_w/a_\eqf) (a_w/a(\eta))^{3(1+w)-4}$.
This ratio, $\rho_\GW/\rho_\rr$, finally becomes $\Omega_\GW$.
For convenience, we define 
\begin{align}
        \tilde \Omega_\GW(\eta,k) \equiv \frac{1}{24} \left( \frac{(1+3w)y}{2} \right)^2 \left( \frac{a_w}{a(\eta)} \right)^{3(1+w)-4} \overline{\mathcal P_h(\eta,k)}.
        \label{eq:til_omega_gw}
\end{align}
Note that this quantity approaches a constant value even during the era with $w \neq 1/3$.
Using this, we can express the GW energy density parameter during the second RD era, which follows the $w$D era, as 
\begin{align}
        \Omega_{\GW,\text{RD}}(\eta\gg \eta_\eqs, k) &\simeq \frac{a_w}{a_\eqf} \tilde \Omega_\GW(\eta_c,k) \nonumber \\
        &\simeq  \frac{\eta_w (\eta_w + 2 \eta_\stf)}{\eta_\stf^2}  \tilde \Omega_\GW(\eta_c,k), 
        \label{eq:omega_til_omega}
\end{align}
where $\eta_c$ is the time when $\tilde \Omega_\GW$ becomes constant. 
More specifically, $\eta_c$ satisfies $\eta_w \ll \eta_c \ll \eta_\eqs$.
The late-time MD era, which begins after the second RD era, dilutes $\Omega_\GW$ and therefore the current energy density parameter of the induced GWs is given by~\cite{Inomata:2020lmk}
\begin{align}
        \Omega_\GW(\eta_0, k) h^2 \simeq 0.39 \left( \frac{g_{*,c}}{106.75} \right) \left( \frac{g_{*s,c}}{106.75} \right)^{-4/3} \Omega_{\rr,0} h^2 \frac{a_w}{a_\eqf} \tilde \Omega_\GW(\eta_c,k),
        \label{eq:omega_gw0}
\end{align}
where $g_{*,c}$ ($g_{*s,c}$) means the effective relativistic degrees of freedom at $\eta_c$ for the energy (entropy) density, $h = H_0/(100 \, \mathrm{km} \, \mathrm{s}^{-1} \, \mathrm{Mpc}^{-1})$ is the reduced Hubble parameter, and $\Omega_{\rr,0}$ is the current radiation energy density parameter ($\Omega_{\rr,0} h^2 \simeq 4.2 \times 10^{-5}$).

We here mention the gauge independence of the induced GWs. 
In general, the second-order tensor perturbations induced by scalar perturbations depend on the gauge~\cite{Matarrese:1997ay,Arroja:2009sh,Hwang:2017oxa,Gong:2019mui,Chang:2020tji,Chang:2020iji,Chang:2020mky}.
During a RD era (as realized after the KD era in our scenario) and in the subhorizon limit, the induced tensor perturbations in generic gauges consist of physical (hence gauge-independent) propagating GWs and gauge artifacts that do not respect the standard dispersion relation of gravitons. 
There are appropriate gauges like the Newtonian gauge, in which the latter contribution can be neglected as the induced tensor perturbations decouple from the scalar sources in the subhorizon limit~\cite{Tomikawa:2019tvi,DeLuca:2019ufz,Inomata:2019yww,Yuan:2019fwv,Lu:2020diy,Domenech:2020xin} (see also Refs.~\cite{Cai:2021jbi,Cai:2021ndu,Ota:2021fdv} for the gauge independence of the GW energy density). 
In other gauges (such as the comoving gauge), the gauge artifact is non-negligible meaning that the induced tensor perturbations do not purely represent physical GWs.
In our analysis, we calculate the GW production until the induced tensor perturbations become propagating GWs in the Newtonian gauge and therefore the final GW spectrum represents the physical (hence gauge-independent) result.

\subsection{Numerical calculation of induced gravitational waves and signal-to-noise ratio}

We here explain how we numerically calculate the induced GWs in our setup with the $w$D era being the KD era ($w=1$). 
Strictly speaking, the transition from the MD era to the KD era is not instantaneous in our setup. 
Throughout this work, we define $\eta_\kk$ as the time at $c_s^2 = 0.95$, after which we can approximately consider that the universe is dominated by the fluid with $w = c_s^2 = 1$.
As seen in the case of $d=1.05$ in Fig.~\ref{fig:phi}, the gravitational potential oscillates and gets suppressed before $\eta_\kk$. This is due to the small but non-negligible $c_s$ before $\eta_\kk$.
Strictly speaking, GWs are also induced before $\eta_\kk$, but the calculation is complicated because the gradual transition of $w$ in the source term and the Green's function should be taken into account during that period. 
In this work, we avoid this complexity and derive conservative GWs by neglecting the contributions before $\eta_\kk$.

To take into account the suppression of the gravitational potential before $\eta_\kk$, we determine $C_{1,2 \bm k}$ by using Eq.~(\ref{eq:c12}) with the values of $\Phi_{\bm k}(\eta_\kk)$ and $\Phi'_{\bm k}(\eta_\kk)$, which are calculated with Eq.~(\ref{eq:phi_eom_w}).
Note that we exactly follow the evolution of $\Phi$ until $\eta_\kk$ without any approximation on the evolution of $w$ and $c_s$.
Using the $C_{1,2 \bm k}$, Eq.~(\ref{eq:trans}), and Eqs.~(\ref{eq:f_def})-(\ref{eq:green}) with $w=1$, we calculate the GW spectrum.
Strictly speaking, we have also neglected the modification of the background evolution due to the non-instantaneous transition in our setup, though its modification should be small given that the timescale of the change of $w$ is shorter than the Hubble timescale in our fiducial parameter sets (see Fig.~\ref{fig:phi}).

For the signal-to-noise ratio, we calculate it by using~\cite{Schmitz:2020syl}
\begin{align}
        \text{SNR} = \left[ n_\text{det} t_\text{obs} \int^{f_\text{max}}_{f_\text{min}} \dd f \left( \frac{\Omega_\text{signal}(f)}{\Omega_\text{noise}(f)} \right)^2 \right]^{1/2},
\end{align}
where $\Omega_\text{signal}$ is the GW signal calculated with Eq.~(\ref{eq:omega_gw0}), $\Omega_\text{noise}$ is the noise spectrum for each observation taken from Ref.~\cite{Schmitz:2020syl}, $t_\text{obs}$ is the observation time, and $n_\text{det}$ distinguishes the cross-correlation measurement ($n_\text{det}= 2$) for DECIGO and BBO and the auto-correlation measurement ($n_\text{det} = 1$) for LISA.
The cutoff frequencies of the integral, $f_\text{max}$ and $f_\text{min}$, are read from the noise spectrum $\Omega_\text{noise}$. 
In Figs.~\ref{fig:ms_fa} and \ref{fig:ma_fa}, we obtain the results by calculating the SNR with the above equation for each point in the figures. 

We note that we do not take into account foreground for the GW observations when we calculate the SNR for Figs.~\ref{fig:ms_fa} and \ref{fig:ma_fa}. 
In particular, the sensitivities of DECIGO and BBO in the region $\lesssim 0.1\,\text{Hz}$ might be reduced by foreground from white dwarf (WD) binaries because the separation of the signals from WD binaries is challenging~\cite{Kawamura:2020pcg}.
We leave the discussion on how much the foreground could change the SNR ratio for future work.

\subsection{Analytic calculation of induced gravitational waves}

We here derive the analytic formula for the induced GW spectrum, Eq.~(\ref{eq:omega_gw_app}) in the main text.
In particular, we present analytic calculations of the integration kernel for the induced GWs, Eqs.~(\ref{eq:calI_J}) and (\ref{eq:calI_Y}), with some approximations and idealization. 
The induced GWs in the KD era were studied in Ref.~\cite{Domenech:2019quo}, but we need to extend their formalism in the presence of multiple cosmological eras. 
We take the instantaneous-transition limit first.  More precisely, we equate the zeroth and first derivatives of $\Phi$ at the transition from the MD era to the KD era. 
After deriving the analytic expression of the induced GWs in the instantaneous transition, we explain how to approximately take into account the effects of the non-instantaneous transition in our fiducial setup.

In the poltergeist mechanism, the dominant contributions originate from deep subhorizon modes of the scalar perturbations.  This means that the argument of the transfer function of $\Phi$ is so large that we can approximate the Bessel functions in the transfer function by their large-argument asymptotics.  
On the other hand, for the Bessel functions in the Green's function of the GWs, we need to separately consider the two cases, $y_w(\equiv k\tilde \eta_w) \gg 1$ and $y_w \ll 1$. 
Note that, in the following, we show the expressions with general $w$ as much as possible for generality, though our main interest lies in the case of $w=1$.

 On small scales with $y_w \gg 1$, the GWs are produced within the horizon. Using the large-argument asymptotics of the Bessel functions, we can approximate Eqs.~(\ref{eq:calI_J}) and (\ref{eq:calI_Y}) in the instantaneous-transition limit as 
\begin{align}
    \mathcal{I}_{J}|_{y_w \gg 1}\simeq&  \sqrt{\frac{2}{\pi}} \frac{3M(x_{\text{eq},1})^2 y_w^{2\beta+1/2}}{50(1+w)} \left( - 2(5+3w) \left( \frac{1}{1-w(t+1)^2} + \frac{1}{1 - w s^2} \right) \sin \left( y_w - \frac{\beta}{2}\pi - \frac{\pi}{4} \right) \right. \nonumber \\
    & \qquad  + 2 (1 + 3w) w u y_w \left( \frac{t+1}{1-w(t+1)^2} + \frac{s}{1-w s^2} \right) \cos \left( y_w - \frac{\beta}{2}\pi - \frac{\pi}{4} \right) \nonumber \\
    &  \qquad  - 2 (1 + 3w) w v y_w \left( \frac{t+1}{1-w(t+1)^2} - \frac{s}{1-w s^2} \right) \cos \left( y_w - \frac{\beta}{2}\pi - \frac{\pi}{4} \right) \nonumber \\
    & \left. \qquad  + (1 + 3w)^2 w u v y_w^2 \left( \frac{1}{1-w(t+1)^2} - \frac{1}{1-w s^2} \right) \sin \left( y_w - \frac{\beta}{2}\pi - \frac{\pi}{4} \right) \right) , \\
    \mathcal{I}_{Y}|_{y_w \gg 1} \simeq&  \sqrt{\frac{2}{\pi}} \frac{3M(x_{\text{eq},1})^2 y_w^{2\beta+1/2}}{50(1+w)} \left(  2(5+3w) \left( \frac{1}{1-w(t+1)^2} + \frac{1}{1 - w s^2} \right) \cos \left( y_w - \frac{\beta}{2}\pi - \frac{\pi}{4} \right) \right. \nonumber \\
    & \qquad  + 2 (1 + 3w) w u y_w \left( \frac{t+1}{1-w(t+1)^2} + \frac{s}{1-w s^2} \right) \sin \left( y_w - \frac{\beta}{2}\pi - \frac{\pi}{4} \right) \nonumber \\
    &  \qquad  - 2 (1 + 3w) w v y_w \left( \frac{t+1}{1-w(t+1)^2} - \frac{s}{1-w s^2} \right) \sin \left( y_w - \frac{\beta}{2}\pi - \frac{\pi}{4} \right) \nonumber \\
    & \left. \qquad - (1 + 3w)^2 w u v y_w^2 \left( \frac{1}{1-w(t+1)^2} - \frac{1}{1-w s^2} \right) \cos \left( y_w - \frac{\beta}{2}\pi - \frac{\pi}{4} \right) \right),
\end{align}
where we have used $t = u+v-1$ and $s=u-v$ instead of $u$ and $v$.
In $k/k_* \ll 1$, where $M(x_\eqf)\simeq 1$, the integral over $t$ is dominated by the large $t(\gg 1)$ region. 
With the large $t$ approximation, the above expressions become
\begin{align}
    \label{eq:calI_J_sub}
    \mathcal{I}_{J}|_{y_w \gg 1} \simeq & -  \sqrt{\frac{2}{\pi}} \frac{3w(1+3w)^2 y_w^{\beta + 5/2}}{200 (1 + w)} \frac{t^2}{1-w s^2}  \sin\left( y_w - \frac{\beta \pi}{2} - \frac{\pi}{4}\right),  \\
    \mathcal{I}_{Y}|_{y_w \gg 1} \simeq & \sqrt{\frac{2}{\pi}} \frac{3w(1+3w)^2 y_w^{\beta + 5/2}}{200 (1 + w)} \frac{t^2}{1-w s^2}  \cos\left( y_w - \frac{\beta \pi}{2} - \frac{\pi}{4}\right) .
\end{align}

On large scales with $y_w \ll 1$, the growth of the tensor mode in the superhorizon regime is also relevant. Corresponding to the superhorizon and subhorizon contributions, we take the small-argument and large-argument asymptotics of the Bessel functions in the Green's function, respectively, and calculate the integrals separately. The sum of these two contributions give the induced GWs on the large scales.  The results for general $w$ involve the incomplete gamma function and are not illuminating, so we focus on the KD case ($w = 1$; $y_w= y_\text{kin}$). 
The large $t$ approximation is always valid for $y_\kk \ll 1$ and we obtain the expression in the instantaneous-transition limit:
\begin{align}
    \mathcal{I}_{J}|_{y_\text{kin} \ll 1} \simeq & \sqrt{\frac{2}{\pi}} \frac{3 y_\kk^3}{50} t^2 \left(-\frac{\sin(1 + s -\pi/4)}{1+s} - \frac{\sin(1-s-\pi/4)}{1-s} + \frac{\sqrt{2\pi} \sin s}{s} \right), \\
    \mathcal{I}_{Y}|_{y_\text{kin} \ll 1}\simeq & \sqrt{\frac{2}{\pi}} \frac{3 y_\kk^3}{50}t^2 \left(\frac{\cos(1 + s -\pi/4)}{1+s} + \frac{\cos(1-s-\pi/4)}{1-s} +6 \sqrt{\frac{2}{\pi}} \frac{(\gamma - \log 2) \sin s - \mathrm{Si}(s)}{s} \right),
    \label{eq:calI_Y_sup}
\end{align}
where $\gamma = 0.577215\dots$ is the Euler-Mascheroni constant and $\mathrm{Si}(z)\equiv \int_0^z \mathrm{d}\bar{z} \, \sin (\bar{z})/\bar{z}$.

Combining the large-scale and small-scale formulas (Eqs.~(\ref{eq:calI_J_sub})-(\ref{eq:calI_Y_sup})), setting $w=1$, and using Eq.~(\ref{eq:p_h_st}), we can obtain the analytic expression of $\mathcal P_h$ for $k/k_* \ll 1$ in the instantaneous-transition limit:
\begin{align}
        \overline{\mathcal P_h(\eta, k)} \simeq&\, 2 \int^\infty_0 \dd t \int^1_{-1} \dd s  \left( \frac{t(2+t)(s^2-1)}{(1-s+t)(1+s+t)} \right)^2 \frac{\pi}{4y} \left( \mathcal{I}_{J}^2 + \mathcal{I}_{Y}^2 \right)  \mathcal P_\zeta(k v) \mathcal P_\zeta(k u)  \nonumber \\
        \simeq&\, 2 \int^{2k_*/k-1}_0 \! \dd t \left(\frac{3}{25}\right)^2 \frac{1}{y} y_\kk^{5} t^4 A^2\times \begin{cases}
    2.14 \times y_\kk & (y_\kk \ll 1) \\
    1 & (y_\kk \gg 1 \text{ and } k/k_* \ll 1)
    \end{cases} \nonumber \\
        =&\,  \frac{576}{3125} \frac{1}{y} y_\kk^{5} \left(\frac{k_*}{k}\right)^5 A^2 \times \begin{cases}
    2.14 \times y_\kk & (y_\kk \ll 1) \\
    1 & (y_\kk \gg 1  \text{ and } k/k_* \ll 1)
    \end{cases},
\end{align}
where we have executed the integral over $s$ (numerically for $y_\kk \ll 1$) assuming the scale-invariant power spectrum given by Eq.~(\ref{eq:pz}).
We have set the upper bound of the $t$-integral to be $2k_*/k-1$ because, even if $k_\tmax > 1/\eta_\eqf$, the contributions on $k > 1/\eta_\eqf$ are automatically suppressed due to the factor $M(k\eta_\eqf)$.
Substituting this into Eq.~(\ref{eq:til_omega_gw}), we obtain 
\begin{align}
        \tilde \Omega_\GW(\eta_c,k) &\simeq \frac{96}{3125} y_\kk^{6} \left(\frac{k_*}{k}\right)^5 A^2 \times \begin{cases}
    2.14 \times y_\kk & (y_\kk \ll 1) \\
    1 & (y_\kk \gg 1  \text{ and } k/k_* \ll 1)
    \end{cases}.
        \label{eq:til_omega_gw_sub}
\end{align}

The effect of the finite duration of the transition from the MD era to the KD era can be approximately taken into account by introducing the normalization factor $Q = \sqrt{T^2(k_*\eta_\kk) + k_*^{-2}{T'}^2(k_*\eta_\kk)}$ (see Ref.~\cite{Inomata:2020lmk} for a similar procedure). 
Multiplying Eq.~(\ref{eq:til_omega_gw_sub}) by $Q^4$ and substituting it into Eq.~(\ref{eq:omega_gw0}) with $y_\kk = k\eta_\kk/4$, we finally obtain Eq.~(\ref{eq:omega_gw_app}) in the main text.

Note that $A$ should be considered to be the amplitude of the power spectrum of curvature perturbations \emph{during} the axion domination. 
Even if the axion field has isocurvature perturbations larger than the curvature perturbations \emph{before} the axion domination, the axion isocurvature contributions are converted to the curvature perturbations during the axion domination.
Because of this, we can safely use the above analytical estimate of the induced GWs for the large axion isocurvature case just by changing $A$ if its power spectrum is almost scale invariant.

We also note that the shape of the induced GW spectrum in our setup is different from that associated with the poltergeist mechanism with the sudden transition from the MD to the RD era~\cite{Inomata:2019ivs}. 
Unlike in the case of the sudden matter-to-radiation transition, there is no resonant amplification around the peak scale in the case of the sudden matter-to-kination transition.
This is because the resonant amplification occurs only when $c_s < 1$, which is required for the induced tensor perturbation and the source scalar perturbations to satisfy the momentum conservation~\cite{Inomata:2019ivs,Domenech:2019quo}. 
Also, the slope of our $\Omega_\text{GW}$ in the range $k \eta_{\text{kin}} \ll 1$ and $k \eta_{\text{eq},2} \gg 1$, which is superhorizon at the beginning of KD era, does \emph{not} satisfy the universal infrared scaling $\propto k^3$~\cite{Cai:2019cdl} because the induced tensor perturbations on those scales continue to be produced until they enter the horizon and become the propagating GWs during the KD era. 
Still, we have confirmed that the induced GWs on the superhorizon scales satisfy the ``causality relation'' $\mathcal{P}_h(k) \propto k^3$ at each time.  The $\Omega_\text{GW} \propto k^2$ scaling is consistent with Ref.~\cite{Domenech:2020kqm}.

\section{Limitation of linear analysis}
\label{smsec:limit_linear}

In this section, we summarize the limitation of the linear analysis from the nonlinearity of the perturbations in the two-field model.

\begin{figure}[t]
  \begin{minipage}[b]{0.49\linewidth}  
    \centering
    \includegraphics[keepaspectratio, scale=0.63]{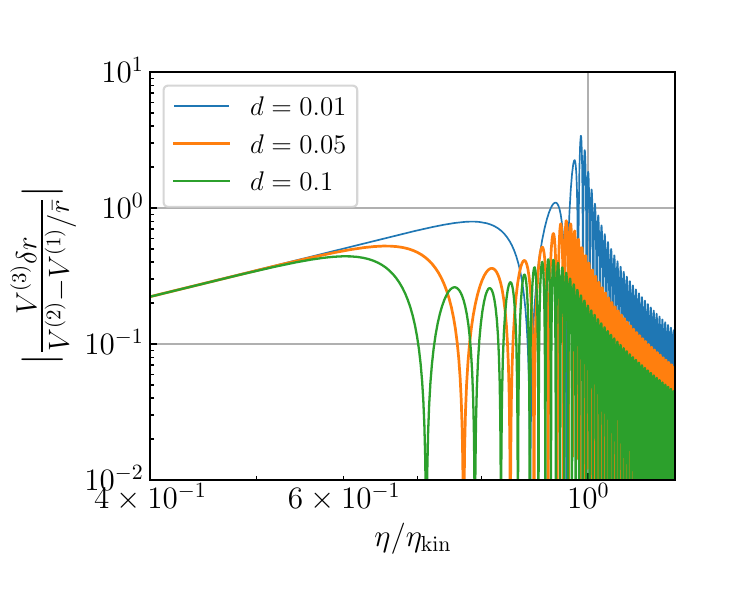}    
  \end{minipage}
  \begin{minipage}[b]{0.49\linewidth}
    \centering
    \mbox{\raisebox{5mm}{\includegraphics[keepaspectratio, scale=0.57]{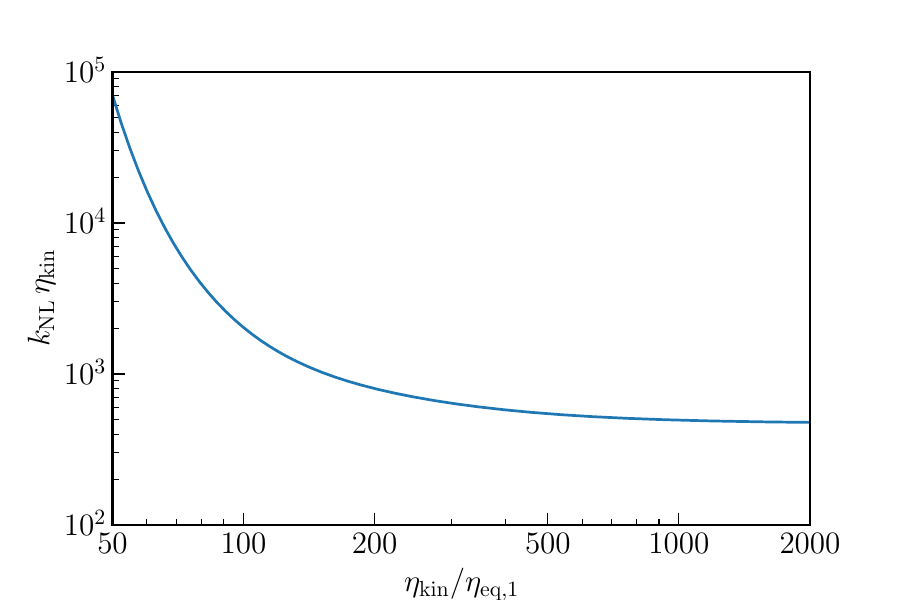}}}
  \end{minipage}
        \caption{
        \emph{Left}:
        The higher-order contribution normalized by the linear order contribution (LHS of Eq.~(\ref{eq:v3_cond})) for $k\eta_\kk = 450$ and $M(k\eta_\eqf) = 1$.
        $\delta r$ is taken to be $\delta r_{\bm k}$ with the replacement of $\zeta_{\bm k} \rightarrow A^{1/2}$.
        \emph{Right}:
         The nonlinear scale and the length of the MD era, based on Eq.~(\ref{eq:k_nl_phi2}).
                $A = 2.1\times 10^{-9}$ is taken in both the figures.
                }  
        \label{fig:nl}
\end{figure}

First, we discuss the nonlinearity in the equation of motion for the radial-field fluctuation.
For convenience, we investigate the nonlinearity with the redefined field $r$, whose potential itself causes the sharp transition of $w$ and $c_s$ with its sharp feature.
From Eq.~(\ref{eq:dS_cond}), $\delta r$ is related to $\delta \theta'$ and $\Phi$ as 
\begin{align}
    \left(a^2  V^\so(\bar r) - (\bar\theta')^2 \right)\delta r \simeq 2 \bar r \left( \bar{\theta }'\delta \theta' - \left(\bar{\theta}'\right)^2 \Phi\right),
    \label{eq:dr_cond}
\end{align}
where we have neglected $\xi k^2/a^2$ because it just makes the correction of $\mathcal O(\xi k^2/(a^2 V^\so))$ in $c_s^2$.
Higher-order potential derivative terms give the corrections to the LHS through $V^{(n)} \delta r^{n-1}$.
If the higher-order contributions are not subdominant, the linear perturbation theory cannot present reliable results.
As a simple diagnostic, we here focus on the lowest-order correction $V^\tho \delta r^2$ and impose 
\begin{align}
    \label{eq:v3_cond0}
    &\left| \frac{V^\tho\delta r}{V^\so - (\bar\theta')^2 }\right| < 1 \\
    \Rightarrow \ &
    \frac{3(\bar r^4 + 16 v^4)}{\bar r^4 - 16 v^4}\left|\frac{\delta r}{\bar r}\right| < 1,
    \label{eq:v3_cond}
\end{align}
where we have substituted Eqs.~(\ref{eq:pot_in_r}) and (\ref{eq:th_v}) to obtain the second line. 
When $d \ll 1$, the matter-to-kination transition occurs near $\bar r \simeq 2v$, which enhances the factor in front of $|\delta r/\bar r|$.
In the case of $\bar r/2v-1 \ll 1$, this condition approximately becomes 
\begin{align}
    \frac{3|\delta r|}{2(\bar r - 2v)} + \mathcal O\left(\left|\frac{\delta r}{\bar r}\right|\right) < 1.
    \label{eq:dr_cond_v3}
\end{align}
For higher-order contributions with $V^{(n)}$, we can similarly obtain 
\begin{align}
    \left|\frac{V^{(n)}\delta r^{n-2}}{V^\so- V^\fo/\bar r} \right| \simeq \mathcal O\left( \frac{|\delta r|^n}{(\bar r - 2v)^{n}} \right).
\end{align}
This indicates that, if Eq.~(\ref{eq:v3_cond}) is satisfied, other higher-order contributions are expected to be subdominant as well.
The left panel of Fig.~\ref{fig:nl} shows the evolution of the LHS of Eq.~(\ref{eq:v3_cond0}), where we here consider $\delta r$ as $\delta r_{\bm k}$ with the replacement of $\zeta_{\bm k} \rightarrow A^{1/2}$.
In this figure, we take $k \eta_\kk = 450$, $A = 2.1\times 10^{-9}$, and $M(k\eta_\eqf) = 1$.
The reason for the choice $k \eta_\kk = 450$ is explained at the end of this section. 
From this figure, we can see that $d = 0.05$ is a marginal case for $A = 2.1\times 10^{-9}$.

Although we have obtained the condition Eq.~(\ref{eq:v3_cond0}) in terms of $r$, we can also obtain a similar condition even in terms of $S$, whose Lagrangian is given by Eq.~(\ref{eq:Leff_in_S}).
Similar to Eq.~(\ref{eq:dr_cond}), the circular mode relation in $S$ is given by (Eq.~(\ref{eq:dS_cond}))
\begin{align}
       &\left(a^2 V^\so - \frac{1}{2}\left(\chi \bar S^2 \right)_{,\bar S\bar S}\left(\bar{\theta }'\right)^2 \right)\delta S 
        \simeq \left(\chi \bar S^2 \right)_{,\bar S} \left(\bar{\theta }'\delta \theta' - \left(\bar{\theta}'\right)^2 \Phi \right),
        \label{eq:dS_cond_lin}
\end{align}
where we have neglected $\xi k^2/a^2$ again.
Unlike in the $r$ basis, the sudden transition in the $S$ basis is caused by the noncanonical form of the axion kinetic term $(\chi S^2) \partial^\mu \theta \partial_\mu \theta$, as mentioned in Sec.~\ref{smsec:two_field_model}.
Because of this, the corrections from the higher order derivative of $(\chi S^2)$ are important. 
Similar to the case in $r$, we here focus on the lowest-order contribution. 
Then, from the LHS of Eq.~(\ref{eq:dS_cond_lin}), the linear perturbation theory requires 
\begin{align}
        &\left|\frac{ \left(a^2V^\tho - \frac{1}{2}\left( \chi\bar S^2 \right)_{,\bar S\bar S\bar S}\left(\bar{\theta }'\right)^2 \right)\delta S}{a^2 V^\so - \frac{1}{2}\left( \chi \bar S^2 \right)_{,\bar S\bar S} \left(\bar{\theta }'\right)^2} \right|< 1 \nonumber \\
        \Rightarrow \ 
        & \left|3\frac{\delta S}{\bar S}\right| < 1,
\end{align}
where we have used Eq.~(\ref{eq:dv_s_int}).
There is no enhancement around $\bar S \simeq \sqrt{2} v$, unlike in the $r$ basis, Eq.~(\ref{eq:v3_cond}).
On the other hand, from the RHS of Eq.~(\ref{eq:dS_cond_lin}), the linear perturbation theory requires
\begin{align}
        &\left|\frac{(\chi \bar S^2)_{,\bar S \bar S}}{(\chi \bar S^2)_{,\bar S}} \delta S\right| < 1 \nonumber \\
        \Rightarrow \ &
        \frac{\bar S^4 + 12 v^4}{\bar S^4 - 4 v^4} \left|\frac{\delta S}{\bar S}\right| < 1.
        \label{eq:non_pertb_cond}
\end{align}
Indeed, we can see that the enhancement around $\bar S \simeq \sqrt{2} v$, which corresponds to the enhancement of the LHS of Eq.~(\ref{eq:v3_cond}) around $\bar r \simeq 2v$.

Next, we discuss the nonlinear scale of the density perturbations.
From Appendix C in Ref.~\cite{Inomata:2020lmk}, the nonlinear scale $k_\NL$ can be related to the amplitude of the curvature power spectrum, given by Eq.~(\ref{eq:pz}) in the main text, as
\begin{align}
  k_\text{NL} (M(k_\text{NL} \eta_{\text{eq},1}))^{1/2}  \mathcal P_\zeta^{1/4}(k_\NL) \simeq \sqrt{\frac{5}{2}} \frac{2}{\eta_\kk},
  \label{eq:k_nl_phi2}
\end{align}
where the density perturbation at $k_\NL$ becomes unity at $\eta_\kk$. 
Note that this equation is based on the assumption that $w=c_s^2 = 0$ in $\eta_\eqf < \eta < \eta_\kk$.
The right panel of Fig.~\ref{fig:nl} shows how the nonlinear scale depends on the length of the MD era, $\eta_\kk/\eta_\eqf$, in the case of $\mathcal P_\zeta(k_\NL)=A$.
We can see that the shorter MD era leads to larger $k_\NL$.
This is because the matter density perturbation only logarithmically grows during the first RD era.
In the limit of long MD era (large $\eta_\kk/\eta_\eqf$ limit), we find $k_\NL \eta_\kk \simeq 467$ for $A = 2.1\times 10^{-9}$. 
This is why we take $k \eta_\kk = 450$ as the benchmark value throughout this work.
For Fig.~\ref{fig:gw} in the main text, we set $k_\tmax = k_\NL$ by using Eq.~(\ref{eq:k_nl_phi2}) to get conservative GW spectra.
Since the transitions of $w$ and $c_s$ in the case of $d = 0.05$ are not instantaneous, the density perturbation at $k_\tmax (\simeq 470/\eta_\kk)$ is $\mathcal{O}(0.1-0.01)$ around the matter-to-kination transition, which can be seen in Fig.~\ref{fig:phi} by using $\delta \propto \Phi$, even if we set $k_\tmax$ with Eq.~(\ref{eq:k_nl_phi2}).
In this sense, our choice of $k_\tmax$ can be considered to be conservative. 
The density perturbation finally becomes $\mathcal O(1)$ around $\eta \simeq \mathcal O(100-1000) \, \eta_\kk$ due to the growth $\delta \propto \eta^{1/2}$ during the KD era.
For Figs.~\ref{fig:ms_fa} and \ref{fig:ma_fa}, we basically set $k_\tmax = k_\NL$ with Eq.~(\ref{eq:k_nl_phi2}), but if we find $k_\tmax > 500/\eta_\kk$, which happens for $k_\tmax \gtrsim 1/\eta_\eqf$, we reset it to $k_\tmax = 500/\eta_\kk$ to reduce the numerical computational cost, which leads to more conservative results.
In particular, this conservative choice of $k_\tmax$ cuts the high-frequency tail of the GW peak conservatively. If we take a larger $k_\tmax$, the observable smallest $m_S$ in Fig.~\ref{fig:ms_fa} would be smaller. Similarly, if the observable smallest $f_a$ in Fig.~\ref{fig:ma_fa} is $f_a > m_S$ (e.g., the case of $m_S = 10^0\,\GeV$), the smallest $f_a$ and the largest $m_a$ would be smaller and larger, respectively, along the $T_\eqs = 2.5\,\MeV$ line.

\end{document}